\documentclass[journal]{IEEEtran}
\usepackage{amssymb}
\usepackage{amsmath}
\usepackage{times}
\usepackage{graphicx}
\usepackage{subfigure}
\usepackage{multirow}
\usepackage{multicol}
\usepackage{amsmath}
\usepackage{bm}
\usepackage{algorithm}
\usepackage{algorithmic}
\usepackage{threeparttable}
\usepackage{dcolumn}
\usepackage{booktabs}
\usepackage{hyperref}
\usepackage{amsthm}
\usepackage{latexsym}
\usepackage{rotating}
\usepackage{textcomp}
\usepackage{colortbl}
\usepackage{enumerate}
\usepackage{cite}
\usepackage{lscape}
\usepackage{color}
\usepackage{epstopdf}
\usepackage[switch]{lineno}
\begin{document}
\title{A Review on Computational Intelligence Techniques\\in Cloud and Edge Computing}
\author{Muhammad Asim, Yong Wang, \emph{Senior Member, IEEE}, Kezhi Wang, \emph{Member, IEEE}, and Pei-Qiu Huang
\thanks{This work was supported in part by the Innovation-Driven Plan in Central South University under Grant 2018CX010, in part by the National Natural Science Foundation of China under Grants 61673397 and 61976225, and in part by the Beijing Advanced Innovation Center for Intelligent Robots and Systems under Grant 2018IRS06. (\emph{Corresponding author: Yong Wang and Kezhi Wang})}
\thanks{M. Asim is with the School of Computer Science and Engineering, Central South University, Changsha 410083, China (Email: asimpk@csu.edu.cn)}
\thanks{Y. Wang and P.-Q. Huang are with the School of Automation, Central South University, Changsha 410083, China (Email: ywang@csu.edu.cn, pqhuang@csu.edu.cn)}
\thanks{K. Wang is with the Department of Computer and Information Sciences, Northumbria University, Newcastle NE1 8ST, UK. (kezhi.wang@northumbria.ac.uk)}}
\maketitle

\begin{abstract}
Cloud computing (CC) is a centralized computing paradigm that accumulates resources centrally and provides these resources to users through Internet. Although CC holds a large number of resources, it may not be acceptable by real-time mobile applications, as it is usually far away from users geographically. On the other hand, edge computing (EC), which distributes resources to the network edge, enjoys increasing popularity in the applications with low-latency and high-reliability requirements. EC provides resources in a decentralized manner, which can respond to users' requirements faster than the normal CC, but with limited computing capacities. As both CC and EC are resource-sensitive, several big issues arise, such as how to conduct job scheduling, resource allocation, and task offloading, which significantly influence the performance of the whole system. To tackle these issues, many optimization problems have been formulated. These optimization problems usually have complex properties, such as non-convexity and NP-hardness, which may not be addressed by the traditional convex optimization-based solutions. Computational intelligence (CI), consisting of a set of nature-inspired computational approaches, recently exhibits great potential in addressing these optimization problems in CC and EC. This paper provides an overview of research problems in CC and EC and recent progresses in addressing them with the help of CI techniques. Informative discussions and future research trends are also presented, with the aim of offering insights to the readers and motivating new research directions.
\end{abstract}


\section*{List of Abbreviations}

\begin{tabular}{l l}
   ACO         & Ant Colony Optimization\\
   AI & Artificial Intelligence\\
   AIS         & Artificial Immune System\\
   BA          & Bee Algorithm\\
   BFA         & Bacterial Foraging Algorithm\\
 CC          & Cloud Computing\\
   CI          & Computational Intelligence\\
   C-RAN       & Cloud-Radio Access Network\\
  CSA         & Cuckoo Search Algorithm\\
\end{tabular}

\begin{tabular}{l l}    DE          & Differential Evolution\\
   EAs         & Evolutionary Algorithms\\
   EC          & Edge Computing\\
   FA          & Firefly Algorithm\\

      	FC        & Fog Computing\\
FI         & Fuzzy Inference\\
   FL          & Fuzzy Logic\\
   FS          & Fuzzy System\\
   GA          & Genetic Algorithm\\
   HS          & Hybrid System\\
   IaaS        & Infrastructure-as-a-Service\\
   IoT         & Internet of Things\\
   LBS         & Learning Based System \\
   MCC         & Mobile Cloud Computing\\
   MDP         & Markov Decision Process\\
   MEC         & Mobile Edge Computing\\
   NN          & Neural Networks\\
   PSO         & Particle Swarm Optimization\\
   QoS         & Quality of Service\\
   SIAs         & Swarm Intelligence Algorithms\\
   \end{tabular}
\section{Introduction}
Cloud computing (CC) is a paradigm of computing technologies that provides on demand services to its clients. The vision of CC is to offer computing, storage, and network resources centrally in the remote clouds, which is related to data centers, backhaul networks, and core networks \cite{M.Ar2, QZhang3}. It is an architecture for allowing appropriate, pervasive, and on request access to a shared pool of configurable resources \cite{mell2011nist}. The large number of resources available in the central cloud can then be leveraged to deliver elastic computing capacity and storage to support resource-constrained end devices. It has been driving the rapid growth of many Internet companies \cite{Yuyi4}. For example, the cloud business has risen to be the most profitable sector for Amazon \cite{N.Wing5}, and Dropbox's success depends highly on the cloud service of Amazon. It can intensify collaboration, scalability, nimbleness, and availability for enterprises as well as users. CC offers services on a pay-as-you-go basis and reduces hardware and software costs, energy, and carbon footprints by providing an optimized and sufficient computing environment \cite{Irena1}. In CC, users first offload tasks to the central cloud, and then the central cloud executes the tasks on behalf of users and sends back results to users \cite{Chen257}. Although CC provides a vast number of resources, easy back and recovery, high accessibility as well as an eco-friendly environment for users (i.e., can be accessed anytime from anywhere on demand), it may not be capable of fulfilling the requirements of real-time applications with low latency and high reliability as the central cloud is far away from users.

\begin{figure*}
	\centering
	\includegraphics[width=.88\textwidth]{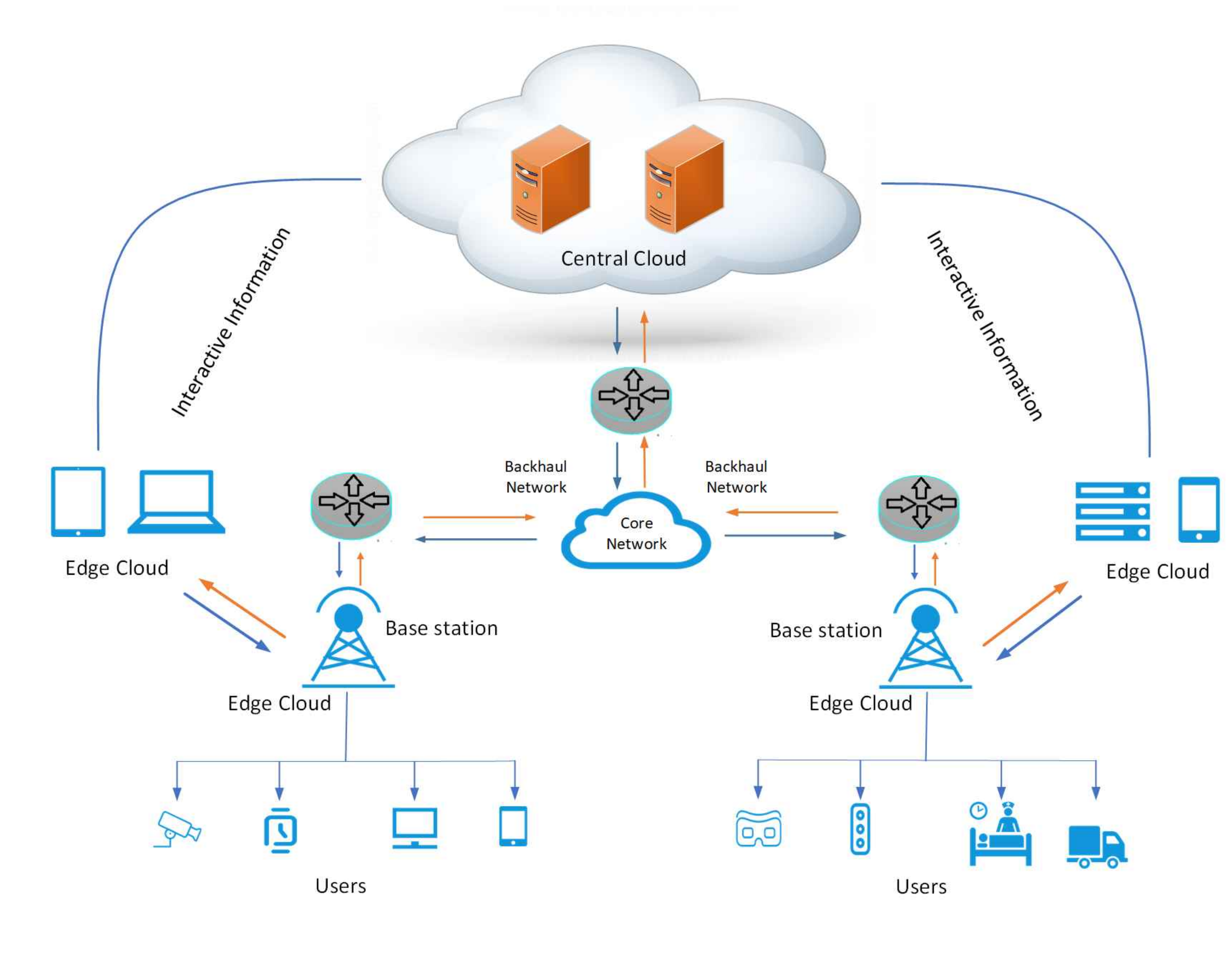}
	\caption{Architecture of CC and EC.}
	\label{FIG-Edge-1}
\end{figure*}

On the other hand, edge computing (EC), which deploys the resources, data, and services from the central cloud to the network edge, enjoys increasing popularity recently. It enables analytics and knowledge generation to occur at or close to users' end. In EC, users first offload tasks to the nearby edge cloud, and then the edge cloud executes their tasks and sends outcomes back to users. Fig. \ref{FIG-Edge-1} presents a typical architecture of CC and EC, where edge clouds can provide computing and network resources between users and the central cloud. Compared with CC, EC can provide services with low latency, high security, and high mobility but with limited computing capacities. CC and EC facilitate a wide range of applications including virtual reality/augmented reality, online game, online video, etc.

In CC and EC, several metrics are related to the quality of service (QoS), such as security and privacy, latency, resource utilization, cost, energy consumption, throughput, and makespan. In order to improve these metrics, we need to consider several issues, such as job scheduling, resource allocation, and task offloading.
For example, security and privacy can be improved by designing efficient task offloading schemes \cite{Yuanfang166}. In addition, latency and energy consumption can be optimized by designing proper resource allocation \cite{Zhaohui154} and task offloading schemes \cite{ZIXUE183, Tang252}. Similarly, other metrics can be improved by considering job scheduling, resource allocation, and task offloading.
To this end, many optimization problems have been formulated for tackling these issues. However, compared with optimization problems in other areas, these optimization problems are normally highly complex and NP-hard, which may include mixed/strongly-coupled variables, nonlinear constraints, multiple objectives, and bilevel structures. Therefore, they may not be solved by the traditional convex optimization-based methods. In addition, their optimal solutions should be found within a reasonable amount of time.

As a class of nature-inspired computational approaches, computational intelligence (CI) exhibits great potential in addressing complex optimization problems, which has attracted much attention from both academia and industry. CI includes evolutionary algorithms (EAs), swarm intelligence algorithms (SIAs), fuzzy system (FS), learning based system (LBS), and hybrid system (HS). These CI techniques have the capability to process imprecise information and search for approximate yet good enough solutions while ensuring robustness and computational tractability \cite{Nicolas9}.

The aim of this paper is to present a review of CI techniques to address issues in CC and EC. Several survey papers and books have been devoted to CC and EC. For example, Hoang \emph{et al.} \cite{Hoang210} presented a survey on architecture, applications, and approaches in mobile CC (MCC). A short review on task scheduling in CC is carried out in \cite{Raja177}. QoS in CC is surveyed in \cite{Helen178} and \cite{Rajeswari180}. Wu \cite{H.Wu199} studied multi-objective decision making for offloading decision in MCC. An extensive survey on MCC is given in \cite{FERNANDO208}. A survey on challenges and opportunities in CC is presented in \cite{Fatemi204}. Zhou \emph{et al.} \cite{Zhou247} published a survey on artificial intelligence (AI) in EC. In \cite{Weisong38}, visions and challenges of EC are discussed. A survey on EC for Internet of things (IoT) is investigated in \cite{WeiYu48}. Liu \emph{et al.} \cite{Liu258} surveyed systems and tools in EC. Peng \emph{et al.} \cite{Kai259} studied service adaptation and provision in mobile EC (MEC). Mao \emph{et al.} \cite{Yuyi4} elaborated on the communication perspective of MEC. Lin \emph{et al.} presented a survey on computation offloading in EC. Wang \emph{et al.} \cite{Jianyu261} focused on offloading algorithms, issues, methods, and perspectives in edge cloud. Moreover, some other surveys devoted to MEC are \cite{Tarik41, Pawani42, Pham206, Pavel39, Arif40, Ren255}. However, none of the aforementioned surveys considers CI techniques. In addition, some survey papers review CI techniques in CC or EC. For example, evolutionary approaches for resource management in CC are reviewed in \cite{Zhan253} and \cite{Mateusz136}. Chopra and Bedi \cite{Pooja181} reviewed the applications of fuzzy logic (FL) in CC. Deep learning (DL) in EC is studied in \cite{Chen246}. It is worth noting that the above-mentioned surveys only consider one particular type of CI for CC or EC.

This survey attempts to give a detailed review of the state-of-the-art CI techniques in CC and EC. Our paper is an ambitious effort to capture the interplay among CI, CC, and EC, instead of delving into one particular CI technique in CC and EC exclusively. The motivation behind this survey is to provide researchers with a glance of mutual relationship between CI and both CC and EC at a higher level.

\begin{figure}
\centering
\includegraphics[width=0.5\textwidth]{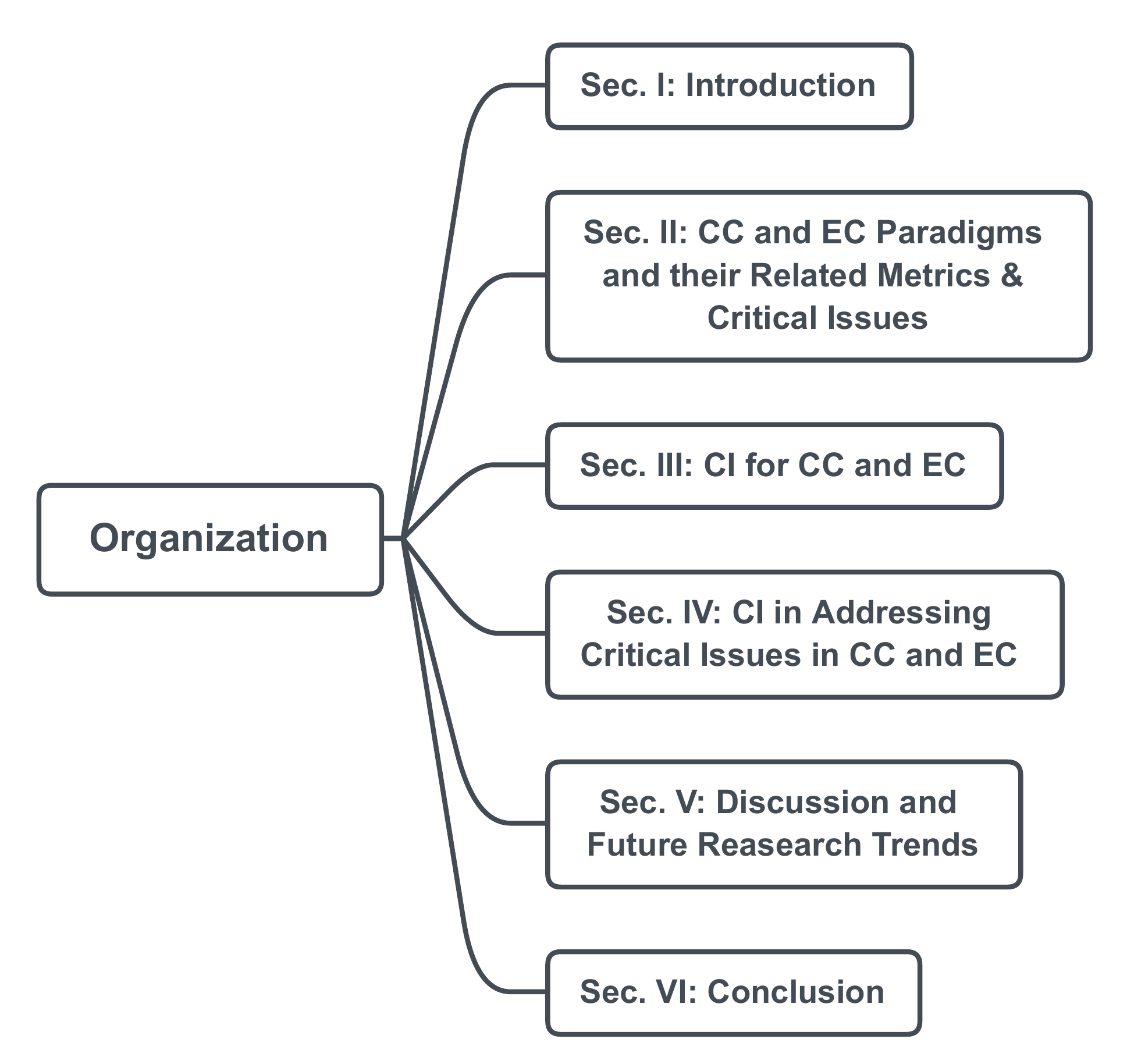}
\caption{Paper Organization.}
\label{FIG-PaperD}
\end{figure}

The works introduced in this survey are taken from relevant journals, workshops, conference proceedings, and theses. This survey focuses on reviewing issues in CC and EC tackled by CI techniques, rather than a detailed study of CC/EC/CI techniques.

The rest of this paper is organized as follows. Section \ref{CC-EC-Paradigms} presents an overview of the concepts related to CC and EC along with important metrics and critical issues. Section \ref{CI-in-CC-EC} introduces CI techniques used for addressing critical issues in CC and EC. Section \ref{App-CI-CC-EC} reviews the existing applications of CI techniques in CC and EC. Finally, Section \ref{Discussion-future} offers a detailed discussion on the use of CI techniques in CC and EC along with future research trends, followed by the conclusion in Section \ref{Conc-CI-CC}. For clarity, the organization of this paper is given in Fig. \ref{FIG-PaperD}.

\section{CC and EC Paradigms and Their Related Metrics and Critical Issues}\label{CC-EC-Paradigms}
\subsection{CC and EC Paradigms}
CC and EC are complementary fields and their features are summarized in Table \ref{CHARACTERICS-Cloud-Edge}. 
\begin{table}
\centering
\caption{Features of CC and EC}
\begin{tabular}{|c|c|c|}
  \hline
  Feature &CC& EC \\
  \hline
  Latency &High& Low \\
  Mobility &No& Yes \\
  Architecture & Centralized & Distributed\\
  Location Awareness &No& Yes \\
  Security & Less Secure & More Secure\\
  Service Access  &Through Core&  At User End\\
  Availability  &High& High   \\
  Geographic Distribution  &No & Yes   \\
  Scalability   &Average& High    \\
  Reliability   & Low   & High \\
   No. of devices connected    &In Millions & In Billions   \\
   Resources & Huge  & Limited \\
  \hline
\end{tabular}
\label{CHARACTERICS-Cloud-Edge}
\end{table}

\subsubsection{CC}
 In CC, different types of paradigms have been designed. Next, we introduce two typical paradigms of CC: MCC and cloud-radio access network (C-RAN).
\begin{itemize}
\item MCC is proposed to enrich resources for users. It merges CC, mobile computing, and wireless networks to enable service providers to help users to conduct computation-intensive tasks \cite{FERNANDO208}. It provides several benefits to users like extending battery lifetime, improving storage capacity and processing power, and providing computation-intensive applications \cite{Hoang210}.

\item C-RAN provides centralized processing, collaborative radio, and energy efficient infrastructure for RAN \cite{Wu212}. In this architecture, network computation-based tasks are performed in the central baseband unit, and the radio signals from distributed antennas are gathered through remote radio heads and transmitted to the cloud by optical transmission network. It offers better services without affecting coverage of network by reducing the number of cell sites and capital expenditures and operating costs.
\end{itemize}
\subsubsection{EC}
EC includes many paradigms, e.g., cloudlet, fog computing (FC), and MEC, which are discussed below.

\begin{itemize}
\item Cloudlet, introduced by Satyanarayanan \emph{et al.} \cite{Satyan63, Satyan64}, is a decentralized and widely distributed infrastructure, i.e., providing storage and processing close to users. It is also referred to as a ``data center in a box''. It is indeed proposed to fulfill the demands of users to reduce response time for new latency-sensitive applications without going to the main central cloud \cite{Gao50}. Although cloudlet has many benefits due to being close to users, it does not have enough computing capacity as the central cloud; thus, it is limited in providing computation-intensive services.

\item FC was first discussed by Cisco in 2012 \cite{F.Bonomi65} and is referred to as an extension of services from the central cloud to the network edge. It reduces the amount of data sent to the central cloud for processing and storage \cite{Chiang295}. It is a paradigm that furnishes computing, storage, and networking services close to users. FC provides better support for real-time applications, dense geographical distribution, low latency, and location awareness.

\item The architecture of MEC was proposed by European Telecommunications Standards Institute in 2014, enabling the provision of resources close to users via RAN \cite{Yi54}. It can provide services with low latency and high bandwidth at the edge of radio network. In addition, MEC makes it possible to measure and improve network performance by setting up services like software defined network and network function virtualization \cite{Hu56}.
\end{itemize}

Although EC can effectively overcome some problems of network congestion and long latency in CC, it has limited computation and communication capabilities \cite{Ren281}. On the other hand, although CC has rich computing resources, it may suffer from high latency. To address the above-mentioned problems, researchers have made some attempts to investigate the hierarchical computing (i.e., the collaboration between CC and EC), such as \cite{Ren281, Shah-Mansouri282, Lan283}, where the tasks of users can be partially or jointly processed at both edge and central clouds.

\subsection{Metrics}
\begin{figure}
\centering
\includegraphics[width=0.5\textwidth]{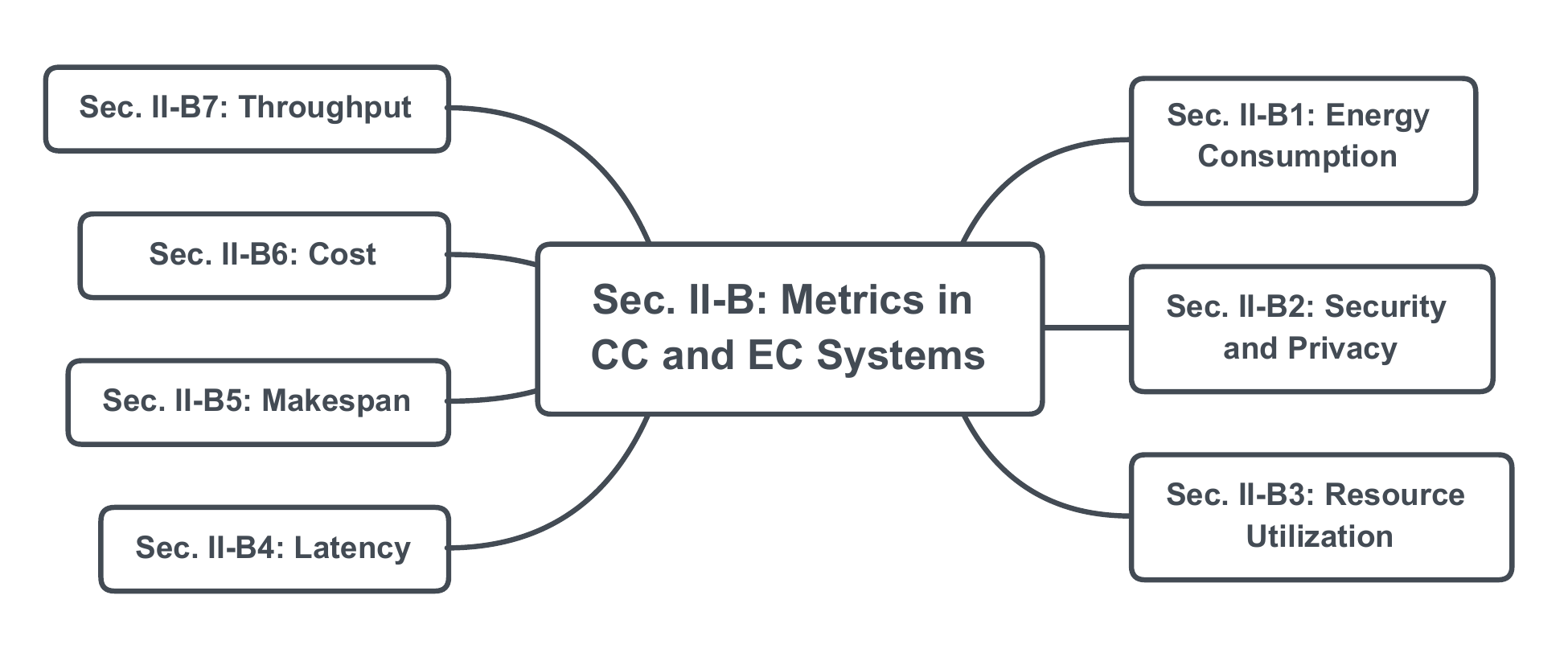}
\caption{Metrics in CC and EC.}
\label{FIG-Metrics}
\end{figure}

Over the past two decades, many researchers focused on diverse aspects of CC and EC, including security and privacy, energy consumption, resource utilization, etc. Some important metrics are given in Fig. \ref{FIG-Metrics} and discussed below.

\subsubsection{Energy Consumption}
 EC may suffer from battery life problem. Even though CC is normally connected to the power grid, energy efficiency is still very important as it is relevant to the profit and economy of CC operators. The energy spent on end devices during task execution is a primary aspect to be considered. Total energy consumption includes static energy consumption and dynamic energy consumption \cite{Ahvar262}. Energy consumed by the system without considering any workload is called the static energy, while the dynamic energy is the energy consumed on the current cloud/edge resources by virtual machines (VMs). Usually, an edge cloud can be a small device like laptop and mobile, and applies the battery. However, CC normally connects to power grid, and has much more power consumption than EC. Therefore, saving energy in EC is to keep edge cloud alive, whereas in CC, saving energy contributes to green society. For users, due to the long distance transmission, CC usually needs more energy than EC for offloading tasks.

\subsubsection{Security and Privacy}
Security and privacy is one of the important metrics to be considered for providing secure and trusted services to users. Many hackers try to interrupt or steal data while users offload and process their tasks at central/edge clouds.  Security and privacy is a set of control-based technologies and policies designed to protect sensitive information, data, applications, and infrastructures \cite{Rameshwar263}. CC is more vulnerable to be attacked by hackers due to the long distance transmission between users and the central cloud. In contrast, EC is a decentralized architecture, which is more secure than CC. Locally sharing, storing, exchanging, and analyzing data among edge clouds make it harder for hackers to get access to data. Moreover, real-time processing and response of EC make it difficult for malicious attackers to detect sensitive information of users. However, EC still inherits many security problems such as privacy leakage, forgery, tampering, spam, jamming, and impersonation \cite{Jianbing46}.

\subsubsection{Resource Utilization}
If CC and EC infrastructures place VMs for resource utilization without any specific scheme, some resources may run out while others are idle. Thus, an efficient resource utilization scheme becomes more important to get the maximum profit of resource utilization by utilizing resources properly \cite{Ala264}. Moreover, load balancing also needs to be considered for better performance and resource utilization. It is essential to distribute workloads across multiple nodes of the system to maintain stability, minimize latency, and improve resource utilization ratio in CC and EC. In CC, proper utilization of resources is needed to save energy, host more users, and make more profits, whereas in EC, it is important to meet high-reliability and low-latency requirements.

\subsubsection{Latency}
A mass of novel mobile applications is emerging, most of them are latency-sensitive. Typically, latency is defined as the delay between a user's request and response of a service provider \cite{Jelassi265}. It has a high effect on the usability and enjoyability of end devices. CC is a large system and hosts a huge number of users, thus leading to routing problems such as VMs' connection problem. In addition, the long physical distance between users and the central cloud also leads to high latency. In contrast, edge clouds are placed at the network edge and thus have a better capability to perform latency-sensitive tasks than CC.

\subsubsection{Makespan}
Makespan is the maximum required time to complete all assigned tasks \cite{Raju266, Mateusz136}, including response time, execution time, waiting time, etc. It can be expressed as:
\begin{equation}\label{PROB3}
     M=\mbox{max} ({T}_i), \,\,\,\,  i = 1,2,\ldots, n
    \end{equation}
where $T_i$ is the completion time of task $i$, and $n$ is the number of tasks.

Compared with EC, there are a huge number of tasks needed to be processed in CC. Therefore, many researchers focused on makespan in CC only.

\subsubsection{Cost}
Since users and service providers usually belong to different entities, users have to pay for availing specific resources \cite{Jelassi265} such as computing or communication resources. The cost can consist of computing cost, running cost, and setup cost. The purpose of this metric is to reduce the cost of services provided to users. One of the most significant ways is to generate a program for services, which can manage the changing behavior of buyers and optimize costs for infrastructure maintenance and order.
\subsubsection{Throughput}
Throughput measures the rate at which data is successfully transferred between two endpoints. It concerns the ability of a service provider to handle the demands of users as it cannot directly track users' experience \cite{Jelassi265}. It is affected by latency and limitations existing at users' devices and central/edge clouds. Usually, EC has a higher throughput compared with CC, as it takes less time in responding to users' requests or transferring data to users.

\subsection{Issues in CC and EC}
In order to improve the above-mentioned metrics in CC and EC, we need to address the following issues.

\subsubsection{Job Scheduling}
The rising demands for services in CC and EC may lead to the imbalanced resource usage and drastically affect the performance of service providers; hence, job scheduling is a necessary prerequisite for improving QoS in CC and EC. The main aim of job scheduling is to order tasks in a specific way. There are two types of job scheduling: static and dynamic scheduling. In static scheduling, all jobs arrive at the same time and are assigned to VMs in a static way, while in dynamic scheduling, all the jobs once arrive, they are scheduled instantly.

\begin{figure}[t]
\centering
\includegraphics[width=0.495\textwidth]{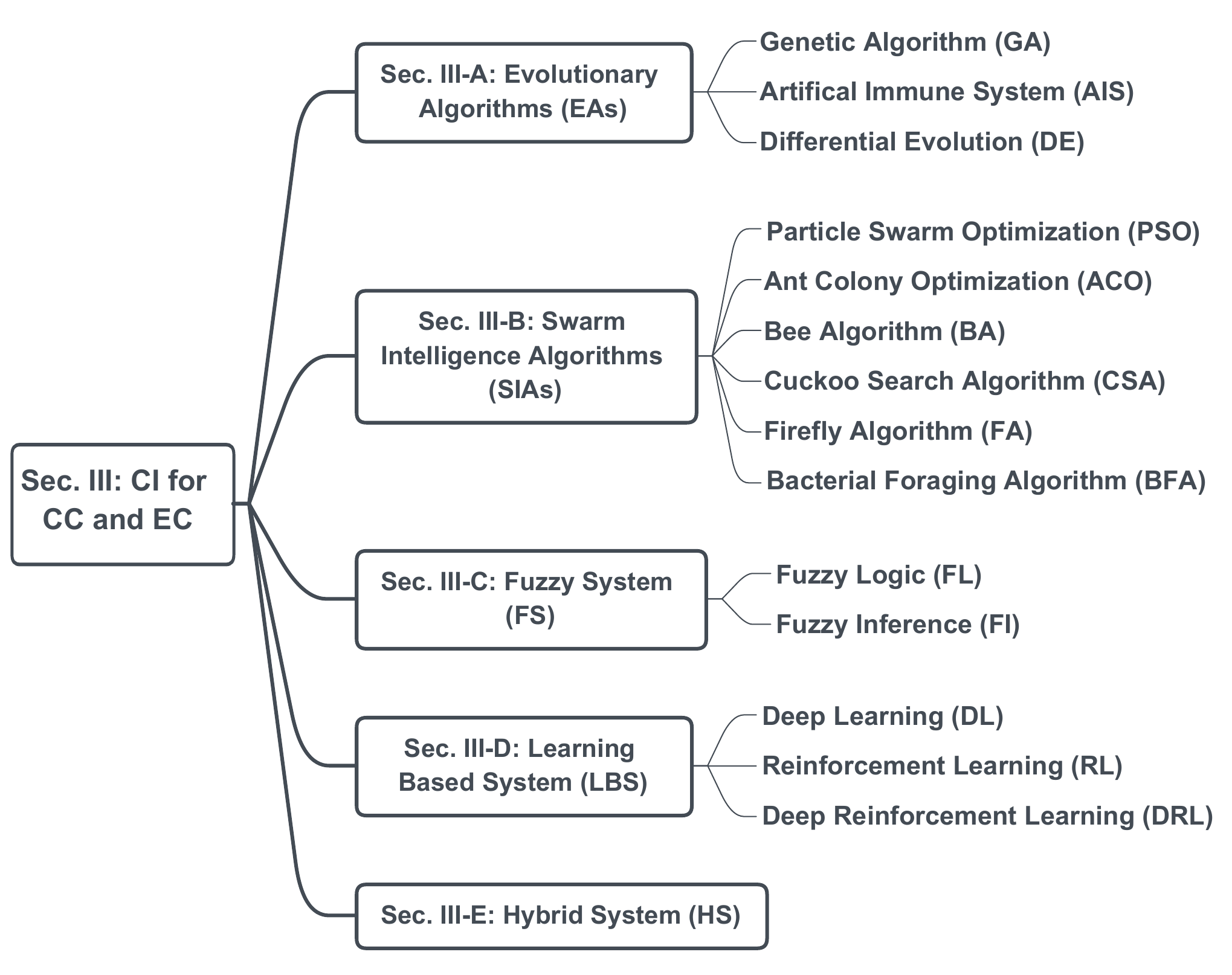}
\caption{Main paradigms of CI family used in CC and EC.}
\label{FIG-CI}
\end{figure}

\subsubsection{Resource Allocation}
Improper distributions of resources in central/edge clouds lead to performance degradation of applications. It is becoming more and more difficult to allocate resources accurately to meet the demands of users and the service level agreements. The main aim of resource allocation is to assign the available resources to the demanded cloud/edge applications over the Internet or wireless networks. Resource allocation should be provided to both users and service providers and, as a result, users can access good-quality services through cloud/edge service providers.

\subsubsection{Task Offloading}
 The rapid growth of Internet services has yielded a variety of computation-intensive applications such as virtual reality applications. If applications are executed in the end devices, it leads to high computing costs, while if they are executed in central/edge clouds, it may take high transmission costs. Thus, users have to decide whether to offload their applications to central/edge clouds or not. Task offloading is the process that maps users' tasks to suitable resources in the form of VMs to execute. Usually, in CC, users only need to consider whether to offload their tasks or not as there is only one data center. But in EC, users have to further consider where to offload their tasks as there are many edge clouds. Task offloading is a critical issue to improve QoS in CC and EC.

\subsubsection{Joint Issues}
 A growing number of applications are multi-sensitive. For instance, some tasks are both resource-intensive and delay-sensitive, e.g., face recognition \cite{Soyata242} and augmented reality applications \cite{Shuwaili243}. Researchers started addressing multiple issues together in CC and EC, which are termed as joint issues.

\section{CI for CC AND EC}\label{CI-in-CC-EC}
CI is a developed paradigm of intelligent systems, which is a set of nature-inspired computational methodologies and approaches to address complex real-world problems. It has attracted much attention from researchers and practitioners around the world. We focus on five categories of CI techniques applied in CC and EC, i.e., EAs, SIAs, FS, LBS, and HS, which are given in Fig. \ref{FIG-CI}.

\subsection{EAs}
EAs try to mimic the process of natural evolution to find suitable solutions to optimization problems \cite{A.Eiben10}. EAs follow evolutionary mechanisms that vary considerably, however the basic ideas of these mechanisms are the same. Some commonly used EAs for CC and EC are given below. 

\begin{itemize}
 \item Genetic algorithm (GA) was proposed by Holland in $1975$ inspired by Darwin's principle of survival of the fittest \cite{JohnHolland14}. GA operates on a set of individuals called a population. Genetic operators, i.e., crossover and mutation, are applied to produce offspring from the parent population. Then, better individuals are selected for the next generation through a selection operator based on the fitness values. This process continues until the stopping criterion is met. Finally, the best individual is returned as the best solution found for an optimization problem. GA has been widely applied to address issues in CC and EC, such as \cite{Hu120, Agarwal225, Jing133, Canali252, Binh245}.

\item Artificial immune system (AIS) is a kind of intelligent computational model inspired by the principles of human immune system with the characteristics of self organization, learning, memory, adaptation, robustness, and scalability \cite{J.Timmis67}. In the initialization phase, candidate antibodies are randomly produced to form a population. The affinity of each antibody is evaluated by the fitness function. In each iteration, each antibody is cloned to a number of offspring. Then, these offspring experience the mutation operator. Only the one with the highest affinity is selected.   Unlike GA, there is no crossover operator in AIS. AIS has been used for addressing issues in CC and EC, such as \cite{Chen170, Rodrigo173, Farhoud174}. 
\item Differential evolution (DE) was developed by Storn and Price in $1997$ \cite{Price82}. Due to the simple structure, ease of implementation, and robustness, it has been successfully applied to complex optimization problems, such as resource allocation \cite{Dolly158} and job scheduling \cite{Wang221} in CC and EC. In DE, the first step is to initialize a population randomly. After initialization, a mutant vector is generated by adding the weighted difference vector of individuals to another individual using a mutation operator. Then, a crossover operator is implemented to obtain a trial vector. Afterward, a selection operator is applied to select better individuals for the next generation. Finally, the best individual is returned as the solution of an optimization problem.

\end{itemize}

\subsection{SIAs}
SIAs are a type of nature-inspired algorithms based on interaction among living organisms \cite{J.Kacp8}. They can be described by the collective behaviors of living organisms working under specific rules. Some commonly used SIAs in CC and EC are discussed in the following:
\begin{itemize}
\item Particle swarm optimization (PSO) was proposed by Kennedy and Eberhart \cite{Kennedy28} in 1995. It is inspired by the social behaviors of the groups of population in nature such as animal herds, bird flocking, or schooling of fish. In PSO, a population of particles is first generated. In each iteration, each particle has an adaptable velocity, according to which it moves in the search space to update position and velocity. Moreover, each particle has a memory, remembering the best position it has ever visited. Thus, its movement is an aggregated acceleration toward its best previously visited position and the best position ever detected/visited by the swarm \cite{K.E.Parsopoulos:35}. This process continues until the stopping criterion is met. PSO has been used to tackle issues in CC \cite{Elina108, Lizheng129, M.Deng191}.

\item Ant colony optimization (ACO), introduced in the early $1990s$ by Dorigo \emph{et al.} \cite{Dorigo59, Dorigo60}, is a technique for optimization. The inspiring source of ACO is the foraging behavior of real ant colonies\cite{Christian57}. In ACO, firstly a finite set of solution components is derived. Secondly, one has to define a set of pheromone values called the pheromone model, which is one of the central components of ACO. In each iteration, ACO assigns higher pheromone values to good solution components probabilistically. ACO has been applied to address job scheduling and resource allocation in CC and EC \cite{Medhat112, Wang229, Shanchen148, Peique185}.
\item Bee algorithm (BA), introduced by Pham\cite{D.T:36} in 2005, is inspired by the natural foraging behavior of honey bees. In BA, the agent population is divided into a small set of scouts and a larger set of foragers. The scouts randomly sample in the solution space, and evaluate the visited flower patches (i.e., locations). In each iteration, BA compares the new solutions discovered by the scouts with the best-so-far solution. The solutions are ordered according to their fitness values, and the highest ranking ones are selected for local search. BA performs exploitative neighborhood search and random explorative search. BA has been applied to job scheduling in CC and EC \cite{Nasrin109, Walaa117, Salim115}. 

\item Cuckoo search algorithm (CSA) is inspired by the living behaviour of a kind of bird named cuckoo and was developed by Yang and Deb in $2009$ \cite{Xin84}. In CSA, the adult cuckoos lay eggs in the habitat of other birds. If these eggs are not found and thrown by host birds, they grow and become mature cuckoos. CSA begins with a primary population i.e., a group of cuckoos. In each iteration, CSA performs the following steps. First, individuals are generated by making modifications in existing individuals. After that, the new individuals are evaluated by the fitness function. The new individuals are compared with existing individuals, if they are better than existing individuals, they replace existing individuals in the population. Job scheduling in CC and EC has been addressed by using CSA \cite{Sup111, Nazir116}. 

\item Firefly algorithm (FA) was developed by Yang \cite{X.S85} in 2008. It is inspired by the flashing pattern and behavior of fireflies. In FA, a population of fireflies is generated randomly. In each iteration, one firefly can be fascinated by other fireflies regardless of its sex. The less brighter firefly is fascinated by the brighter one in the population. The brightness of a firefly is set on the landscape of the fitness function \cite{X.S87}. In case there is no brighter one, the fireflies move randomly and form a new population for the next iteration. The light intensity of fireflies is updated by evaluating the new population. Resource allocation and job scheduling in CC and EC have been tackled by using FA \cite{Nidhi157,Amanpreet152, Kanza163}. 

\item Bacterial foraging algorithm (BFA) is a swarm optimization algorithm inspired by colonies of escherichia coli and proposed by Passino \cite{Passino244}. In BFA, individuals are first generated randomly. Then, for all individuals, the chemotaxis process is carried out. The process in which an organism moves with the gradient of a substance concentration is called chemotaxis. After the chemotaxis process, individuals are reproduced and dispersed. Elimination-dispersion is applied as a final step, in which many random individuals are eliminated and many new individuals are generated. BFA has been used for job scheduling in CC \cite{Juhi105, Liji145}.

\end{itemize}
\subsection{FS}
FS is a type of CI family that resembles human reasoning. FS is a unique system that can handle numerical data and linguistic knowledge at the same time. We have found two types of FS for CC and EC, which are described below.
\begin{itemize}
\item FL was initiated by Zadeh in $1965$ \cite{L.A88} during the development of theory of fuzzy sets. It works on approximate reasoning rather than exact approximation. The main difference between FL and traditional logic is that FL represents natural language and human thinking. In traditional logic, each member of a binary set has a logical value either 0 or 1, which shows that either it fully belongs to the set or not. There is no concept of partial membership in the traditional set theory \cite{L.A89}. Fuzzy modeling provides the best alternative for fuzziness and obscurities. FL has been applied to address issues in CC and EC, for instance, job scheduling \cite{Anindita107, Haratian215, Srinivas123}, resource allocation \cite{Xiaojun217, Haratian215}, and task offloading \cite{Ritu167}.
\item Fuzzy inference (FI) is a model that can infer a crisp value as an input from a set of inputs, their fuzzy set, membership function, and a set of inference rules \cite{T.Takagi90}. FI is usually constructed as follows. Firstly, inputs for the model are fuzzified through a membership function linked with the predefined fuzzy set. Linguistics terms are used to label the fuzzy set. Afterward, the firing strength is calculated for each rule, showing that some rules are more important than the others in approaching the conclusion. All firing strengths are then aggregated and weighted for all rules, thus producing a fuzzy value. Finally, this fuzzified value is defuzzified by using a proper method. FI has been used for resource allocation \cite{Tao216} and task offloading \cite{Guanwen168, Chenhao171} in CC and EC. 
\end{itemize}

\begin{table*}[t]
	\centering
	\caption{EAs for Addressing Issues in CC and EC.}
	\begin{center}
		\begin{tabular}{|c|c|c|c|c|c|}
			\hline
			\multicolumn{1}{|c|} {Issue}& {EA}& {Metric}&{Paradigm} &{Reference}   \\  \cline{1-5}
			\multicolumn{1}{|c|}{\multirow{7}{*}{Job Scheduling}} &
			\multicolumn{1}{c|}  {GA} & {execution time}&{CC} & {\cite{Hu120}}   \\ \cline{2-5}
			\multicolumn{1}{|c|}{}                            &
			\multicolumn{1}{c|}   {GA} & {response time}&{CC} & {\cite{Agarwal225}}   \\ \cline{2-5}
			\multicolumn{1}{|c|}{}                            &
			\multicolumn{1}{c|}   {GA} & {energy consumption, cost}&{CC} & {\cite{Jing133}}  \\\cline{2-5}
			\multicolumn{1}{|c|}{} &
			\multicolumn{1}{c|}  {GA} & {makespan}&{CC} &{\cite{Zarina118}} \\ \cline{2-5}
			\multicolumn{1}{|c|}{} &
			
			\multicolumn{1}{c|}   {GA} & {makespan, resource utilization}&{CC} & {\cite{Pardeep119}} \\ \cline{2-5}
			\multicolumn{1}{|c|}{} &
			\multicolumn{1}{c|}   {GA} & {makespan, computing cost}&{CC} & {\cite{Shaminder132}}  \\ \cline{2-5}
			\multicolumn{1}{|c|}{}                            &
			\multicolumn{1}{c|}   {GA} & {execution time, cost}&{FC} & {\cite{Binh245}}  \\ \cline{2-5}
			\multicolumn{1}{|c|}{}                            &
			\multicolumn{1}{c|}   {AIS} & {response time}& {FC} & {\cite{Wang250}}  \\ \cline{2-5}
			\hline
			\multicolumn{1}{|c|}{\multirow{3}{*}{Resource Allocation}} &
			\multicolumn{1}{c|}  {DE} & {resource utilization, makespan}&{CC} &{\cite{Dolly158}}  \\ \cline{2-5}
			
			\multicolumn{1}{|c|}{} &
			\multicolumn{1}{c|}  {GA} &{makespan} &{CC}&  {\cite{Yang164}}   \\ \cline{2-5}
			\multicolumn{1}{|c|}{}                            &
			\multicolumn{1}{c|}   {GA} &{energy consumption, latency} & {MEC}&{\cite{Zhaohui154}} \\ 
			\hline
			\multicolumn{1}{|c|}{\multirow{10}{*}{Task Offloading}} &
			\multicolumn{1}{c|}  {AIS} & {security and privacy}&{CC} &{\cite{Chen170}}  \\ \cline{2-5}
			\multicolumn{1}{|c|}{} &
			\multicolumn{1}{c|}  {GA} &{execution time, energy consumption}&{MCC} &  {\cite{S.Deng190}}  \\ \cline{2-5}
			\multicolumn{1}{|c|}{} &
			\multicolumn{1}{c|}  {GA} & {energy consumption} &{MCC} &{\cite{Nitesh156}} \\ \cline{2-5}
			\multicolumn{1}{|c|}{} &
			\multicolumn{1}{c|}  {GA} &{latency} &{FC}&  {\cite{Canali252}}  \\ \cline{2-5}
			\multicolumn{1}{|c|}{} &
			\multicolumn{1}{c|}  {AIS} &{security and privacy}&{EC} &  {\cite{Rodrigo173}}   \\ \cline{2-5}
			\multicolumn{1}{|c|}{} &
			\multicolumn{1}{c|}  {AIS} & {security and privacy}&{FC} &{\cite{Farhoud174}}    \\ \cline{2-5}
			\multicolumn{1}{|c|}{} &
			\multicolumn{1}{c|}  {GA} & {latency}&{EC} &{\cite{ZIXUE183}} \\ \cline{2-5}
			\multicolumn{1}{|c|}{} &
			\multicolumn{1}{c|}  {GA} & {energy consumption}&{MEC} &{\cite{Tang252}}   \\ \cline{2-5}
			\multicolumn{1}{|c|}{} &
			\multicolumn{1}{c|}  {GA} & {execution time, energy consumption}&{MCC} &{\cite{Goudarzi276}}   \\ \cline{2-5}
			\multicolumn{1}{|c|}{} &
			\multicolumn{1}{c|} {GA} & { energy consumption, latency}&{MEC} &{\cite{Bozorgchenani292}}   \\ 
			\hline
			\multicolumn{1}{|c|}{\multirow{3}{*}{Joint issues}} &
			\multicolumn{1}{c|}   {DE} & {energy consumption}&{MEC} & {\cite{Wang221}} \\ \cline{2-5}
			\multicolumn{1}{|c|}{} &
			\multicolumn{1}{c|}  {GA} &{energy consumption} &{MEC}&  {\cite{Guo251}} \\ \cline{2-5}
			\multicolumn{1}{|c|}{} &
			\multicolumn{1}{c|} {GA} & {makespan} & {MEC}&  {\cite{Li277} }\\ 
			\hline
		\end{tabular}
		\label{tab-EAs}
	\end{center}
\end{table*}

\subsection{LBS}
 LBS is inspired by the learning behaviour of living organisms. It uncovers the relationship between the set of nominal features and states or objects \cite{I.H93}. In the following, we present the three most prominent types of LBS.
\begin{itemize}
\item DL is inspired by human's thinking ability, i.e., the mechanisms of human brain and neurons for processing signals, and was proposed by Hinton \emph{et al.} \cite{Hinton201} in 2006. It can cataract layers to extract features from the input data and eventually form a decision \cite{Q.Mao200}. DL consists of input layer, output layer, and one or more hidden layers. In the input layer, the state of environment is represented in a suitable numerical form, which can be used as inputs of the network. Afterward, an activation function (e.g., Sigmoid function, Tanh function, or rectified linear unit function) is applied to represent or interrupt the recollection of results. Subsequently, all weights among layers are updated. Finally, the structure of DL is constructed. DL has been used for task offloading in EC \cite{Yuanfang166, Huang238, Rani218}. 

\item Reinforcement learning (RL) \cite{R.S94, Y.Shohan95} determines an optimal policy dictating which actions to take at certain states to achieve the highest possible reward. The problem such as resource allocation \cite{Almuthanna159} in CC and EC is often formulated as a Markov decision process (MDP), where there is a set of states and actions. In RL, an agent under a given state selects a suitable action from the set of actions, receives a reward, and moves into a new state where it chooses another action from the updated set of actions. This process continues until a specific stopping criterion is met. RL has been applied to address issues in CC and EC \cite{Wang254, Zhiping138, Almuthanna159, Xavier162, liu220, Dinh213}. 

\item Deep RL (DRL) can be seen as a class of new efficient learning algorithm by combining DL with RL \cite{Youssef284, Mosavi285, Arulkumaran286}. It is a powerful model that implements DL architectures such as deep neural networks with RL algorithms like Q-learning to scale and solve problems in different areas \cite{Lavet296}. It has been effectively used for addressing issues in CC and EC, for instance, job scheduling \cite{Jianpeng137, Qingchen155}, resource allocation \cite{Yang219}, task offloading \cite{Meng214, Xianfu188, Zhang224, Ning241}, and joint issues \cite{LiangHuang182, Liu222, Ning223}. Two commonly used DRL algorithms are deep Q-learning and deep Q-network.
\end{itemize}

\subsection{HS}
HS is a system combining two or more CI techniques to overcome their individual shortcomings. For example, genetic-fuzzy system is proposed for automating maritime risk assessment \cite{A.Teske97}, which is a hybrid system of GA and FL. Also, EAs and SIAs are often combined as both have similar behaviors. Similarly, fuzzy adaptive theory \cite{S.G98} is sometimes hybridized with FL and neural networks (NN) \cite{Nicolas9}. Recently, some hybrid algorithms have been applied to address critical issues in CC \cite{Ahmad124, Liu226, RASHIDI230, N.Moganarangan192}.

\section{CI in Addressing Critical Issues in CC and EC}\label{App-CI-CC-EC}
This section reviews the CI techniques introduced in Section \ref{CI-in-CC-EC} to solve the critical issues in CC and EC: job scheduling, resource allocation, task offloading, and joint issues.

\subsection{EAs in CC and EC}
This subsection is relevant to the works on the use of EAs in CC and EC, which are summarized in Table \ref{tab-EAs}.

\subsubsection{EAs for Job Scheduling}
Hu \emph{et al.} \cite{Hu120} proposed an improved adaptive GA based on a priority mechanism for task scheduling in CC. This algorithm ensures the least execution time for job scheduling and guarantees the QoS requirements of users. It outperforms an adaptive GA and some other GAs in terms of convergence speed, feasibility and effectiveness.
Agarwal and Srivastava \cite{Agarwal225} proposed a GA for task scheduling in CC, which distributes the loads among VMs effectively to optimize the overall response time. Experimental results show that the proposed algorithm outperforms some existing techniques such as greedy-based techniques and First Come First Serve in terms of overall response time.
Liu \emph{et al.} \cite{Jing133} developed a job scheduling model in CC based on multi-objective GA, which aims to minimize the energy consumption and maximize the profit of service. The proposed model has several components to analyze the applications and to allocate the appropriate resources to the applications. Experimental results show that the proposed model can obtain a higher profit, while consuming lower energy. 
Zarina \emph{et al.} \cite{Zarina118} investigated GA-based optimal job scheduling and load balancing in CC. They combined a GA with three traditional scheduling techniques: min-min, max-min and suffrage. In the first phase, these three traditional scheduling techniques are applied to obtain the minimum completion time for a given job at each VM. After that, GA is applied to attain better QoS by utilizing available resources. Experimental results reveal that the proposed algorithm outperforms min-min, max-min, suffrage, and First Come First Serve in terms of makespan.
Kumar and Verma \cite{Pardeep119} presented an improved GA by using min-min and max-min techniques in the original GA for scheduling tasks in CC. In this improved GA, the initial population is generated by using min-min and max-min techniques. Makespan is considered as the fitness function. This improved GA is applied for two cases. In the first case, the number of VMs is kept constant and the number of cloudlets is varied, while in the second case, the number of cloudlets is fixed and the number of VMs is varied. Experimental results show that it performs better than the simple GA in terms of makespan and resource utilization.
Shaminder \emph{et al.} \cite{Shaminder132} proposed a modified GA by merging two existing scheduling algorithms with standard GA for scheduling tasks in CC. In the proposed algorithm, the initial population is generated by using the output schedules of two algorithms (i.e., longest cloudlet to fastest processor and smallest cloudlet to fastest processor) and $8$ random schedules. To achieve time minimization and compare it with existing heuristics, a fitness function is formulated for single-user jobs. Experimental results show that the proposed algorithm performs better than standard GA in terms of response time.

Binh \emph{et al.} \cite{Binh245} proposed a GA-based algorithm for solving task scheduling in cloud-FC. The proposed algorithm tries to achieve the optimal tradeoff between the execution time and operating costs by addressing different tasks in cloud-FC. Experimental results demonstrate that the proposed algorithm outperforms BA in terms of time-cost tradeoff.
Wang \emph{et al.} \cite{Wang250} studied an immune scheduling network-based method for task scheduling in FC. The proposed method uses forward and backward propagation in the ad hoc network along with the power of distributed schedulers to generate the optimized scheduler strategies to deal with computing node overloaded and achieve the optimal task finishing time reducing. Experimental results reveal that the proposed method performs better than the modified critical path, dynamic critical path, dominant sequence clustering, and GA.

\subsubsection{EAs for Resource Allocation}
Dolly \cite{Dolly158} used DE for resource allocation in CC. The proposed method avoids premature convergence effectively, reduces the makespan, and increases the resource utilization. The proposed method outperforms PSO in terms of makespan, resource utilization, and load balancing.
Lin and Zhong \cite{Yang164} presented a GA-based strategy for computing resource allocation in CC. They incorporated enhancements and local search into GA for solving computing resource allocation in CC. The computation time is focused in this work, which is an important factor in cloud manufacturing for fast response to users' requests. The proposed method is compared with other algorithms like PSO, BA, ACO, etc. Experimental results show the effectiveness of the proposed method in terms of makespan.
Luo \emph{et al.} \cite{Zhaohui154} proposed a GA-based cashing placement strategy to minimize energy consumption in MEC. A joint optimization problem is formulated by considering energy consumption, backhaul capacities, and content popularity distributions. A GA is applied to solve this complicated joint optimization problem. Simulation results show that the proposed algorithm effectively determines the near-optimal caching placement and obtains better performance in terms of energy efficiency compared with the conventional caching placement strategies.

\subsubsection{EAs for Task Offloading}
Chen and Yang \cite{Chen170} presented a data security strategy based on AIS in CC. They discussed the main factors affecting data security, introduced a Hadoop distributed file system, and applied AIS with negative selection and dynamic selection in CC. Simulation results show that the proposed strategy performs better than GA, PSO, and simulated annealing.
Deng \emph{et al.} \cite{S.Deng190} proposed a GA-based computation task offloading approach for MCC. This approach is designed for robust offloading decision to optimize execution time and energy consumption of mobile services. Numerical results demonstrate that with near-linear algorithmic complexity, the proposed approach can produce near-optimal solutions.
Kaushik and Kumar \cite{Nitesh156} applied GA to a framework designed for elastic applications to reduce communication and computation energy by finding the optimum offloading solution. They focused on augmented execution of mobile applications and formulated an optimization problem by minimizing the cost function, which is the combination of communication energy and computation energy. Simulation results demonstrate that the proposed approach outperforms all mobile-side execution and all cloud-side execution.

\begin{table*}[htbp]
	\centering
	\caption{SIAs for Addressing Issues in CC and EC.}
	\begin{center}
		\begin{tabular}{cc|c|c|c|c|}
			\hline
			\multicolumn{1}{|c|} {Issue}& {SIA}& {Metric}&{Paradigm} &{Reference}   \\  \cline{1-5}
			\multicolumn{1}{|c|}{\multirow{17}{*}{Job Scheduling}} &
			
			\multicolumn{1}{c|} {PSO} & {throughput, response time}&{CC} & {\cite{Elina108}}   \\ \cline{2-5}
			\multicolumn{1}{|c|}{}                            &
			\multicolumn{1}{c|} {PSO} & {makespan, cost}&{CC} & {\cite{Lizheng129}}     \\ \cline{2-5}
			\multicolumn{1}{|c|}{}                            &
			\multicolumn{1}{c|} {ACO} & {makespan} &{CC}& {\cite{Medhat112}}    \\ \cline{2-5}
			\multicolumn{1}{|c|}{}                            &
			\multicolumn{1}{c|} {ACO} & {energy consumption, resource utilization} &{MCC} & {\cite{Wang229}}    \\ \cline{2-5}
			\multicolumn{1}{|c|}{}                            &
			\multicolumn{1}{c|} {ACO} & {resource utilization}&{CC} & {\cite{Zehua130}}   \\ \cline{2-5}
			\multicolumn{1}{|c|}{}                            &
			\multicolumn{1}{c|} {ACO} & {energy consumption} &{MCC} & {\cite{Wei231}}   \\ \cline{2-5}
			\multicolumn{1}{|c|}{}                            &	
			\multicolumn{1}{c|} {BA} & {makespan}&{CC} & {\cite{Nasrin109}}    \\ \cline{2-5}
			\multicolumn{1}{|c|}{}                            &
			\multicolumn{1}{c|} {BA} & {response time, execution time}&{CC} & {\cite{Walaa117}}    \\ \cline{2-5}
			\multicolumn{1}{|c|}{}                            &	
			\multicolumn{1}{c|} {CSA} & {setup cost, running cost}&{CC} & {\cite{Sup111}}    \\ \cline{2-5}
			
			\multicolumn{1}{|c|}{}                            &
			
			\multicolumn{1}{c|} {FA} & {energy consumption, resource utilization} &{CC} & {\cite{Nidhi157}}    \\ \cline{2-5}
			\multicolumn{1}{|c|}{}                            &
			\multicolumn{1}{c|} {FA} & {makespan}&{CC} & {\cite{Amanpreet152}}  \\ \cline{2-5}
			\multicolumn{1}{|c|}{}                            &
			\multicolumn{1}{c|} {BFA} & {makespan, computation cost, resource utilization}&{CC} & {\cite{Juhi105}}  \\ \cline{2-5}
			\multicolumn{1}{|c|}{}                            &
			\multicolumn{1}{c|} {BFA} & {cost, makespan}&{CC} & {\cite{Liji145}}   \\ \cline{2-5}
			
			\multicolumn{1}{|c|}{}                            &
			\multicolumn{1}{c|} {BA} & {execution time}&{FC} & {\cite{Salim115}}    \\ \cline{2-5}
			\multicolumn{1}{|c|}{}                            &
			\multicolumn{1}{c|} {CSA} & {response time, cost}&{CC \& FC} & {\cite{Nazir116}}    \\ \cline{2-5}
			\multicolumn{1}{|c|}{} &
			\multicolumn{1}{c|} {ACO} & {energy consumption, waiting time}&{CC}  & {\cite{Shanchen148}}  \\ \cline{2-5}
			\multicolumn{1}{|c|}{}  &
			\multicolumn{1}{c|} {ACO} & {energy consumption, resource utilization}&{CC} & {\cite{X.F.LIU193}}  \\ 
			\hline
			\multicolumn{1}{|c|}{\multirow{1}{*}{Resource Allocation}} &
	
			\multicolumn{1}{c|} {FA} & {resource utilization, cost}&{CC \& FC} & {\cite{Kanza163}}   \\
			
			\hline
			\multicolumn{1}{|c|}{\multirow{2}{*}{Task Offloading}} &
			\multicolumn{1}{c|} {PSO} & {energy consumption, latency} & {CC} & {\cite{M.Deng191}}   \\ \cline{2-5}
			\multicolumn{1}{|c|}{} &
			\multicolumn{1}{c|} {ACO} & {response time}&{MCC} & {\cite{Bao227}}  \\ 
			\hline
			\multicolumn{1}{|c|}{\multirow{1}{*}{Joint Issues}} &
			\multicolumn{1}{c|} {ACO} & {energy consumption}& {MEC} & {\cite{Peique185}} \\
			\hline
		\end{tabular}
		\label{tab-SIAs}
	\end{center}
\end{table*}

Canali \emph{et al.} \cite{Canali252} proposed a GA for service placement in FC. They studied mapping data streams over fog nodes and presented an optimization model. Then, a scalable heuristic based on GA is adopted to tackle this model. Experimental results verify the stability and performance of the proposed heuristic. 
Roman \emph{et al.} \cite{Rodrigo173} proposed an AIS-based strategy for the IoT system using edge technologies. In the proposed strategy, the requirements of immune system for IoT are analyzed and a security architecture is proposed to meet the demands of users. It makes a decision on the number and type of VMs deployed at the edge infrastructure.
Farhoud \emph{et al.} \cite{Farhoud174} proposed a new distributed and lightweight intrusion detection system in FC based on an AIS model. In this paper, the intrusion detection system is distributed to three layers: cloud, fog, and edge layers. Intrusion detection system, primary network traffic clustering, and detectors' training are performed in the cloud layer. Intrusion alerts are analyzed by using smart data concept in the fog layer. Detectors are deployed in edge clouds in the edge layer. Experimental results show the efficiency of the proposed system.
Cheng \emph{et al.} \cite{ZIXUE183} proposed a fast heuristic algorithm based on GA for just-in-time code offloading for wearable computing. The proposed algorithm is compared with offloading nothing, offloading all to cloud, and simple greedy offloading. Numerical results show that the proposed algorithm outperforms the three competitors significantly.
Tang \emph{et al.} \cite{Tang252} suggested a task caching and migration strategy in MEC based on GA to satisfy the completion time constraints and the goal of minimizing energy consumption. They adopted a fine-grained task partitioning migration model to transform users' tasks into directed graphs with multiple subtasks. The task caching is proposed to further reduce the delay and energy consumption. Simulation results reveal that the proposed strategy can greatly reduce the energy consumption of end devices compared with all tasks being executed in the edge or end devices. 
Goudarzi \emph{et al.} \cite{Goudarzi276} proposed a GA-based algorithm for multi-site computation offloading in MCC. They modified genetic operators to reduce ineffective solutions and evaluated the efficiency of the proposed algorithm using graphs of real mobile applications. Experimental results demonstrate that the proposed algorithm outperforms other existing algorithms in terms of execution time, energy consumption, and weighted cost in a timely manner.
Bozorgchenani \emph{et al.} \cite{Bozorgchenani292} proposed a multi-objective EA, i.e., non-dominated sorting GA, to find the optimal offloading decisions in MEC. They modeled the task offloading  as a constrained multi-objective optimization problem that jointly minimizes the task processing delay and the energy consumption of mobile devices. Experimental results reveal that the proposed approach outperforms other existing approaches in terms of energy consumption and task processing delay.

\subsubsection{EAs for Joint Issues}
Wang \emph{et al.} \cite{Wang221} studied a multi-unmanned aerial vehicle (UAV) enabled MEC, where the delay-sensitive task can be executed on the local device or one of UAVs. This paper designs a two-layer optimization method for jointly optimizing the deployment of UAVs and task scheduling, where DE and the greedy method are adopted at the upper layer and the lower layer, respectively.
Guo \emph{et al.} \cite{Guo251} presented a computation offloading model in the multi-access and multi-channel interference MEC. The offloading decision, channel allocation, and computation resource allocation are formulated as a mixed-integer nonlinear programming problem. A suboptimal algorithm, i.e., a GA-based computation algorithm, is proposed to solve this large-scale and NP-hard problem. Simulation results demonstrate that the proposed algorithm outperforms a PSO-based computation algorithm, a random computation algorithm, and a local computation algorithm in terms of energy consumption.
Li and Zhu \cite{Li277} proposed a joint optimization method based on GA to optimize offloading proportion, channel bandwidth, and mobile edge servers' computing resource allocation in MEC. Simulation results demonstrate that the proposed algorithm can effectively reduce the task completion time and guarantee fairness among users.

\subsection{SIAs in CC and EC}
This subsection discusses the works on the use of SIAs for solving issues in CC and EC, as summarized in Table \ref{tab-SIAs}.

\subsubsection{SIAs for Job Scheduling}
Elina \emph{et al.} \cite{Elina108} described a two-level cloud scheduler by using PSO, which operates under the IaaS model. They explored whether the use of a priority-based policy at the VM-level is suitable in an online cloud. Experimental results show that the proposed algorithm offers a good balance between throughput and response time.
Guo \emph{et al.} \cite{Lizheng129} proposed a PSO algorithm by incorporating a small position value rule to minimize processing cost in CC. Experimental results show that the proposed algorithm outperforms two other PSO-based algorithms in terms of convergence, processing time, and processing cost.
Medhat \emph{et al.} \cite{Medhat112} presented an ACO algorithm for task scheduling in CC. At first, parameters with better values are determined for ACO through experiments. Then, ACO is evaluated on a set of instances with up to 1000 tasks.
Wang \emph{et al.} \cite{Wang229} investigated MCC-assisted execution of a multi-task scheduling problem in a hybrid MCC architecture, and formulated it as an optimization problem with time constraints. Cooperative multi-task scheduling based on ACO is used to solve this optimization problem by considering task profit, task deadline, task dependence, node heterogeneity, and load balancing.
Zhang and Zhang \cite{Zehua130} applied an ACO algorithm for load balancing in a complex network named ACCLB for CC. The purpose of this study is to cope with the dynamic load balancing problem in open CC federation. Experimental results show that the proposed method performs better than SearchMax-SearchMin and classic ACO in terms of minimizing standard deviation and gains a more suitable distribution of the workloads on the whole cloud federation.
Wei \emph{et al.} \cite{Wei231} studied application scheduling in MCC with load balancing by using ACO. They presented a hybrid local mobile cloud model by extending the cloudlet architecture. The proposed model can select applications with the maximum profit and minimum energy consumption in a heavy load environment. 

Hesabian \emph{et al.} \cite{Nasrin109} presented a scheduling method in CC based on BA to dedicate the sources optimally. The proposed method behaves like a load balancing BA for small-scale systems in terms of makespan, while outperforms it in large-scale systems. 
Walaa \emph{et al.} \cite{Walaa117} proposed a BA for load balancing in CC, which distributes workloads of multiple network links to avoid under utilization and over utilization of the resources. The proposed algorithm can minimize the response time and data center processing time by distributing workloads on different VMs based on the availability and load of each VM. It performs better than modified throttled and round robin algorithms in terms of average response time and execution time.
Supacheep \emph{et al.} \cite{Sup111} applied CSA for job scheduling in CC. They considered only static cases of job scheduling. The results of CSA are better than those of GA in most cases and it can find larger number of feasible solutions than GA.
Kansal and Chana \cite{Nidhi157} proposed an energy-aware VM migration technique based on FA for CC, which migrates the maximally loaded VM to the least loaded active node while maintaining the performance and energy efficiency of the data centers. This technique is compared with ACO-based and FFD-based techniques in the Cloudsim simulator.
Kaur and Sherma \cite{Amanpreet152} applied FA to address workflow scheduling in CC, which reduces the execution time by efficient scheduling of workflow. The proposed FA performs better than other SIAs for workflow scheduling in terms of makespan. 
Verma \emph{et al.} \cite{Juhi105} proposed an improved BFA for scheduling tasks in CC. The performance of this algorithm is evaluated by using CloudSim toolkit. Experimental results demonstrate that it outperforms BFA and two other algorithms in terms of makespan, computation cost, and resource utilization.
Jacob \emph{et al.} \cite{Liji145} presented a BFA for resource scheduling in CC. In this work, a hyper-heuristic-based scheduling algorithm is used in the cloud system to map the resources in an efficient way. The proposed BFA reduces the cost and makespan and outperforms some existing algorithms like GA, ACO, a priority-based algorithm, and a Berger model-based algorithm.

\begin{table*}[t]
	\centering
	\caption{FS for Addressing Issues in CC and EC.}
	\begin{center}
		\begin{tabular}{|c|c|c|c|c|}
			\hline
			\multicolumn{1}{|c|} {Issue}& {FS}& {Metric}&{Paradigm} &{Reference}    \\  
			\hline
			\multicolumn{1}{|c|}{} &
			\multicolumn{1}{c|}  {FL} & {response time, throughput, resource utilization}&{CC} &{\cite{Anindita107}}  \\ \cline{2-5}
			\multicolumn{1}{|c|}{\multirow{3}{*}{Job Scheduling}} &
			\multicolumn{1}{c|}   {FL} & {latency}&{CC} & {\cite{Srinivas123}}  \\ \cline{2-5}
			\multicolumn{1}{|c|}{} &
			\multicolumn{1}{c|}  {FL} & {latency, cost}&{CC} & {\cite{Awatif122}}   \\ \cline{2-5}
			\multicolumn{1}{|c|}{}                            &
			
			\multicolumn{1}{c|}  {FL} & {latency}&{CC} & {\cite{Sally128}} \\ \cline{2-5}
			\multicolumn{1}{|c|}{} &
			\multicolumn{1}{c|}  {FL} & {resource utilization} &{CC}& {\cite{Himadri147}}  \\ 
			\hline
			\multicolumn{1}{|c|}{\multirow{3}{*}{Resource Allocation}} &
			\multicolumn{1}{c|}  {FL} & {resource utilization, cost} & {CC}& {\cite{Haratian215}}   \\ \cline{2-5}
			\multicolumn{1}{|c|}{} &
			\multicolumn{1}{c|}  {FL} & {cost}&{CC}  & {\cite{Xiaojun217}} \\ \cline{2-5}
			\multicolumn{1}{|c|}{}                            &
			\multicolumn{1}{c|}   {FI} & {resource utilization}&{CC} & {\cite{Tao216}} \\ 
			
			\hline
			\multicolumn{1}{|c|}{\multirow{3}{*}{Task Offloading}} &
			\multicolumn{1}{c|}  {FL} & {security and privacy}&{CC} &{\cite{Ritu167}}  \\ \cline{2-5}
			\multicolumn{1}{|c|}{} &
			\multicolumn{1}{c|}  {FI} & {security and privacy}&{CC} & {\cite{Chenhao171}}   \\ \cline{2-5}
			\multicolumn{1}{|c|}{} &
			\multicolumn{1}{c|}  {FI} & {security and privacy, execution time}&{MEC} & {\cite{Guanwen168}} \\
			\hline
			\multicolumn{1}{|c|}{\multirow{2}{*}{Joint Issues}} &
			\multicolumn{1}{c|}  {FL} & {resource utilization, response time}&{EC \& CC} &{\cite{Sonmez288}}  \\ \cline{2-5}
			\multicolumn{1}{|c|}{} &
			\multicolumn{1}{c|}  {FL} & {response time, processing time}&{ CC} &{\cite{Zulkar289}}  \\
			
			\hline
		\end{tabular}
		\label{tab-FS}
	\end{center}
\end{table*}

Bitam \emph{et al.} \cite{Salim115} proposed a BA for job scheduling in FC, in which two factors are considered: the CPU execution time and the allocated memory required by the overall tasks. The reliability and efficiency of the proposed BA are evaluated by performing a set of tests. It is superior to GA and PSO in terms of allocated memory and execution time.
Nazir \emph{et al.} \cite{Nazir116} proposed a CSA for balancing load of users' requests in residential areas in cloud-FC. The performance of CSA is compared with that of some existing techniques like round robin and throttled algorithms.
Pang \emph{et al.} \cite{Shanchen148} designed an ACO algorithm to address dynamic energy management in cloud data center. They used Petri net for analyzing scheduling process, and then a task-oriented resource allocation method is proposed to optimize the running time and energy consumption of the system.
Liu \emph{et al.} \cite{X.F.LIU193} proposed an algorithm called OEMAC based on ACO to minimize energy for VM placement in CC. OEMAC minimizes the number of active servers. Experimental results show that OEMAC outperforms conventional heuristics and some other EAs in terms of saving energy, improving resource utilization, minimizing the number of active servers, and balancing different resources.
\subsubsection{SIAs for Resource Allocation}
Kanza \emph{et al.} \cite{Kanza163} utilized a FA for efficient resource allocation in cloud-FC, in which load balancing and cost reduction are considered.

\subsubsection{SIAs for Task Offloading}
Deng \emph{et al.} \cite{M.Deng191} adopted discrete PSO to search the optimal offloading policy in cloud-enhanced small cell networks, with the aim of minimizing energy consumption under strict delay constraints. The energy-efficient task offloading problem is formulated as a constrained 0-1 programming problem. Experimental results show that the proposed approach outperforms local execution and conventional rough-granularity offloading policy in terms of energy saving.
Bao \emph{et al.} \cite{Bao227} proposed an ACO-based method for addressing computation offloading problem in MCC. Different rules are adopted to provide services to a large number of service requests.

\subsubsection{SIAs for Joint Issues}
Huang \emph{et al.} \cite{Peique185} studied an ACO-based bilevel optimization approach for joint offloading decision and resource allocation in cooperative MEC to attain energy-efficient task execution under delay constraints. The effectiveness of the proposed approach is demonstrated by comparing it with four other algorithms.

\begin{table*}[t]
	\centering
	\caption{LBS for Addressing Issues in CC and EC.}
	\begin{center}
		\begin{tabular}{|c|c|c|c|c|}
			\hline
			\multicolumn{1}{|c|} {Issue}& {LBS}& {Metric}&{Paradigm} &{Reference}    \\  
			\hline
			\multicolumn{1}{|c|}{\multirow{4}{*}{Job Scheduling}} &
			\multicolumn{1}{c|}  {RL} & {makespan}&{CC} & {\cite{Zhiping138}}   \\ \cline{2-5}
			\multicolumn{1}{|c|}{} &
			\multicolumn{1}{c|}  {DRL} & {latency}&{CC} & {\cite{Jianpeng137}}  \\ \cline{2-5}
			\multicolumn{1}{|c|}{} &			
			\multicolumn{1}{c|}  {DRL} & {energy consumption} &{MEC} & {\cite{Qingchen155}} \\ 
			\hline
			\multicolumn{1}{|c|}{\multirow{7}{*}{Resource Allocation}} &
			\multicolumn{1}{c|}  {RL} & {response time, resource utilization}&{CC} & {\cite{Xavier162}}\\ \cline{2-5}			
			\multicolumn{1}{|c|}{} &
			\multicolumn{1}{c|}  {RL} & {latency} &{FC}& {\cite{Almuthanna159}}  \\ \cline{2-5}
			\multicolumn{1}{|c|}{} &			
			\multicolumn{1}{c|}  {RL} & {energy consumption, latency}&{EC} & {\cite{liu220}}  \\ \cline{2-5}
			\multicolumn{1}{|c|}{} &			
			\multicolumn{1}{c|}  {DRL} & {latency}&{MEC} & {\cite{Yang219}}  \\ \cline{2-5}
			\multicolumn{1}{|c|}{} &
			\multicolumn{1}{c|}  {DL} & {response time, resource utilization}&{EC} & {\cite{Nguyen160}}  \\ \cline{2-5}
			\multicolumn{1}{|c|}{} &
			\multicolumn{1}{c|}   {DRL} &  {response time}& {MEC} &  {\cite{Wang278}}  \\ \cline{2-5}
			\multicolumn{1}{|c|}{} &
			\multicolumn{1}{c|}   {DRL} &  {resource utilization, makespan}& {MEC} &  {\cite{Xiong291}}\\
			\hline
			\multicolumn{1}{|c|}{\multirow{13}{*}{Task Offloading}} &
			\multicolumn{1}{c|}  {DRL} & {resource utilization, latency}&{MCC} & {\cite{Li186}}  \\ \cline{2-5}
			\multicolumn{1}{|c|}{} &			
			\multicolumn{1}{c|}  {RL} & {resource utilization, latency}& {MCC} & {\cite{Sowndarya189}}  \\ \cline{2-5}
			\multicolumn{1}{|c|}{} &
			\multicolumn{1}{c|}  {DL} & {security and privacy} & {MEC} & {\cite{Yuanfang166}} \\ \cline{2-5}
			\multicolumn{1}{|c|}{} &
			\multicolumn{1}{c|}  {DL} & {energy consumption, makespan} & {Cloudlet} & {\cite{Rani218}} \\ \cline{2-5}
			\multicolumn{1}{|c|}{} &
			\multicolumn{1}{c|}  {RL} & {energy consumption} & {MEC} & {\cite{Dinh213}} \\ \cline{2-5}
			\multicolumn{1}{|c|}{} &
			\multicolumn{1}{c|}  {DRL} & {energy consumption, latency}& {MEC} & {\cite{Meng214}} \\ \cline{2-5}
			\multicolumn{1}{|c|}{} &
			\multicolumn{1}{c|}  {DRL} &{latency, cost}& {MEC} & {\cite{Xianfu188}}   \\ \cline{2-5}			
			\multicolumn{1}{|c|}{} &
			\multicolumn{1}{c|}  {RL} & {processing time, latency}&{MEC} & {\cite{Zhang224}}   \\ \cline{2-5}
			\multicolumn{1}{|c|}{} &
			\multicolumn{1}{c|}  {DRL} & {energy consumption}&{FC} & {\cite{Ning241}}  \\ \cline{2-5}
			\multicolumn{1}{|c|}{} &
			\multicolumn{1}{c|}  {RL} & {energy consumption}&{MEC} & {\cite{Shermila187}}   \\ \cline{2-5}
			\multicolumn{1}{|c|}{} &
			\multicolumn{1}{c|}  {DRL} &  {latency}& {MEC} & {\cite{Wang272}}   \\ \cline{2-5}
			\multicolumn{1}{|c|}{} &
			\multicolumn{1}{c|} {DRL} &  {latency, energy consumption}& {MEC} &  {\cite{Chen275} }  \\ \cline{2-5}
			\multicolumn{1}{|c|}{} &
			\multicolumn{1}{c|}  {DRL} &  {latency, energy consumption, computation cost}& {MEC} &  {\cite{Chen279}}  \\
			\hline
			\multicolumn{1}{|c|}{\multirow{7}{*}{Joint Issues}} &
			\multicolumn{1}{c|}  {DL} & {energy consumption, cost}&{MEC} & {\cite{Huang238}}  \\ \cline{2-5}
			\multicolumn{1}{|c|}{} &
			\multicolumn{1}{c|}  {DRL} & {cost, energy consumption, latency}&{MEC} & {\cite{LiangHuang182}} \\ \cline{2-5}			
			\multicolumn{1}{|c|}{} &
			\multicolumn{1}{c|}  {DRL} & {latency, resource utilization}&{EC} & \cite{Liu222}  \\ \cline{2-5}
			\multicolumn{1}{|c|}{} &
			\multicolumn{1}{c|}  {DRL} & {energy consumption, execution time}&{EC} & \cite{Ning223}  \\ \cline{2-5}
			\multicolumn{1}{|c|}{} &
			\multicolumn{1}{c|}  {RL} & {security and privacy, energy consumption}&{MEC} & {\cite{Liang169}} \\ \cline{2-5}
			\multicolumn{1}{|c|}{} &
			\multicolumn{1}{c|} {DRL} &  {latency, energy consumption}& {MEC} &  {\cite{Li273}} \\ \cline{2-5}
			\multicolumn{1}{|c|}{} &
			\multicolumn{1}{c|}  {DRL} &  {latency}& {MEC} &  {\cite{Huang274}}\\ 
			\hline
		\end{tabular}
		\label{tab-LS}
	\end{center}
\end{table*}

\subsection{FS in CC and EC}
The works devoted to the use of FS in CC and EC are discussed in this subsection and summarized in Table \ref{tab-FS}.

\subsubsection{FS for Job Scheduling}
Anindita \cite{Anindita107} proposed a new dynamic task scheduling approach in CC based on FL, which takes two inputs: the time required to complete the tasks in each VM and the number of requests received from each VM. These inputs produce an output i.e., id of VM which is assigned to the host. Experimental results show that the proposed algorithm outperforms First Come First Serve, round robin, and BA in terms of throughput, response time, and resource utilization.
Srinivas \emph{et al.} \cite{Srinivas123} proposed an efficient load balancing algorithm by using FL based on round robin load balancing technique for attaining better resource utilization in CC. For evaluating the balanced load through FL, two parameters are used: processor speed and assigned load of a VM. Numerical results reveal that the proposed algorithm outperforms round robin load balancing technique in terms of minimizing the processing and response time.
Ragmani \emph{et al.} \cite{Awatif122} proposed an improved scheduling strategy based on FL in CC to evaluate processing time, response time, and total cost. FL uses the number of VMs, bandwidth, the number of processors, processor speed, the size of data, and request per user per hour as inputs. Then, the fuzzy controller calculates the global performance indicator. 
Issawi \emph{et al.} \cite{Sally128} proposed an efficient adaptive load balancing algorithm in CC based on FL and round robin/random assignment, which consists of three main components: burst detector, load balancing algorithm, and fuzzifier. When a request is received in the data center, it is detected by the burst detector as normal or burst workload state. Then, a load balancing algorithm (round robin in burst and random assignment in non-burst state) is selected, which assigns the received task to an appropriate VM based on the information provided by the fuzzifier. Experimental results show that the proposed algorithm reduces the response and processing time.
Mondal \emph{et al.} \cite{Himadri147} used FL to improve QoS in CC by balancing the imposed load in the system. This paper considers the speed of processor and load as inputs and load balancing as output. 

\subsubsection{FS for Resource Allocation}
Haratian \emph{et al.} \cite{Haratian215} proposed an adaptive and fuzzy resource management approach called AFRM for allocating resources in CC. In AFRM, the last resource values of each VM are collected through the environment sensors and sent to a fuzzy controller. Then, AFRM analyzes the received information to make a decision on how to reallocate the resources in each iteration of a self-adaptive control cycle. Experimental results show that AFRM outperforms rule-based and static-fuzzy approaches in terms of resource allocation efficiency, utility, the service level agreement violations, and cost.
Wang \emph{et al.} \cite{Xiaojun217} proposed a strategy for resource allocation based on improved fuzzy clustering in CC. They first divided the set of resources in CC into different resource pools on the service level of users by using improved fuzzy clustering, and then generated a resource scheduling scheme by using a scheduling algorithm. The proposed strategy shows certain advantages in terms of the number of iterations and the accuracy of classification compared with other algorithms.
Tao \emph{et al.} \cite{Tao216} designed a containerized test environment in CC and developed a node selection algorithm based on FI for container deployment, where FI is applied to dynamically predict the most proper node for the deployment of the selected containers. Experimental results show that the proposed algorithm performs better than existing container deployment algorithms.

\subsubsection{FS for Task Offloading}
Ritu and Jain \cite{Ritu167} presented a trust model in CC by using FL. The proposed model makes use of turnaround time, availability, and reliability for evaluating trust in CC.
Qu and Buyya \cite{Chenhao171} proposed a cloud trust evaluation system by using hierarchical FI for service selection in CC. To facilitate service selection, this system evaluates the trust of clouds according to users' fuzzy QoS requirements and services' dynamic performance. Simulations and case studies demonstrate the effectiveness and efficiency of this system.
Li \emph{et al.} \cite{Guanwen168} designed an architecture in MEC to decouple the security functions with physical resources and developed a FI-based algorithm to find the optimal order of the required security functions. Numerical results show that the proposed algorithm performs better than a widely used fuzzy-based simple additive weighting algorithm in terms of inverted generational distance values and execution time.

\subsubsection{FS for Joint Issues}
Sonmez \emph{et al.} \cite{Sonmez288} proposed an FL-based approach for workload orchestration in EC, where execution locations for incoming tasks from mobile devices are decided within an EC infrastructure. Simulation results demonstrate that the proposed approach performs better than other algorithms for the cases studied in terms of resource utilization and response time.
Zulkar \emph{et al.} \cite{Zulkar289} presented a FL-based dynamic load balancing algorithm in virtualized data centers, which can efficiently predict the VM where the next job is scheduled. They modeled the requirements of memory, bandwidth, and disk space using FL. Simulation results demonstrate that the proposed algorithm outperforms other scheduling algorithms in terms of response and processing time in the data centers.

\subsection{LBS in CC and EC}
In this subsection, we review the works on the use of LBS in CC and EC as summarized in Table \ref{tab-LS}.

\subsubsection{LBS for Job Scheduling}
Peng \emph{et al.} \cite{Zhiping138} proposed a RL-based mixed job scheduler scheme for CC, which considers accurate scaled CC environment and efficient job scheduling under VM resource and service level agreement constraints. The proposed scheme outperforms fast-fit, best-fit, min-min, and max-min scheduling schemes in terms of makespan.
Lin \emph{et al.} \cite{Jianpeng137} presented a multi-resource cloud job scheduling strategy in CC based on current popular DRL and deep Q-network, which aims to reduce the average job completion time and average job slowdown. In this strategy, the convolutional NN is adopted to perceive the system resources, job state features, and RL decision-making capabilities to solve online awareness decision problems in CC. Based on the experimental results, the proposed strategy performs better than classical heuristic algorithms and converges faster than a standard policy gradient algorithm.
Zhang \emph{et al.} \cite{Qingchen155} combined a stacked auto-encoder with a Q-learning model to design a deep Q-learning model for energy-efficient scheduling in real-time systems. The main function of the stacked auto-encoder is to replace the Q-function for learning the Q-value. The proposed model can save 4.2\% energy compared with hybrid dynamic voltage and frequency scaling scheduling based on Q-learning for different sets of tasks.

\subsubsection{LBS for Resource Allocation}
Dutreilh \emph{et al.} \cite{Xavier162} studied RL for autonomic resource allocation in CC, in which proper initialization is adopted at the early stages and convergence speedups are applied in the learning phases.
Nassar and Yilmaz \cite{Almuthanna159} proposed a RL-based resource allocation algorithm in fog radio access networks. They formulated the resource allocation problem as an MDP in two alternative formulations: infinite-horizon MDP and finite-horizon MDP. Experimental results show that the proposed algorithm outperforms the fixed-threshold method.
Liu \emph{et al.} \cite{liu220} proposed a RL approach based on $\epsilon$-greedy Q-learning for resource allocation in EC. Experimental results show the effectiveness of the proposed approach in terms of minimizing energy consumption and latency.
Yang \emph{et al.} \cite{Yang219} proposed DRL-based resource allocation in MEC. They allocated computation resources by investigating different strategies in MEC networks that operate with finite block length codes to support low-latency communications. Simulation results show that the proposed algorithm outperforms the random and equal scheduling benchmarks. 
Luong \emph{et al.} \cite{Nguyen160} proposed an optimal auction based on DL in the edge resource allocation. They constructed a multi-layer NN architecture based on analytical solution of the optimal auction. The NN first performs monotone transformations of the miners' bids. Then, allocation and conditional payment rules are calculated for the miners. Simulation results show that the proposed scheme can quickly converge to a solution.
Wang \emph{et al.} \cite{Wang278} proposed a DRL-based approach for smart resource allocations including computing resource allocation and network resource allocation in MEC. They considered two aspects: average service time minimization and resource allocation balancing. Experimental results reveal that the proposed  approach outperforms the traditional open shortest path first algorithm.
Xiong  \emph{et al.} \cite{Xiong291} proposed a DRL-based approach for resource allocation of IoT in EC. They formulated the resource allocation problem as an MDP. They also proposed an improved deep Q-network algorithm for policy learning, where multiple replay memories are applied to separately store the experiences with small mutual influence. Simulation results show that the proposed algorithm outperforms the original deep Q-network algorithm in terms of convergence, and that the corresponding policy performs better than other policies regarding completion time.

\subsubsection{LBS for Task Offloading}
Quan \emph{et al.} \cite{Li186} proposed a two-layered RL algorithm for task offloading with a tradeoff between physical machine utilization rate and delay in MCC. The k-nearest neighbors algorithm divides the physical machines into many clusters. The first layer selects a cluster via learning the optimal policy, while the second layer learns an optimal policy to choose the optimal physical machine to execute the current offloaded task. Numerical results show that the proposed algorithm is faster than DRL when learning the optimal policy for task offloading.
Sundar and Liang \cite{Sowndarya189} proposed a game and learning approach for multi-user computation offloading in MCC. This paper discusses both online and offline computation task offloading from multiple users to a cloud or nearby cloud at the edge. The offline algorithm provides a better average solution while online algorithm is much faster.

\begin{table*}[t]
	\centering
	\caption{HS for addressing issues in CC and EC.}
	\begin{center}
		\begin{tabular}{|c|c|c|c|c|c|}
			\hline
			\multicolumn{1}{|c|} {Issue}& {HS}& {Metric} &{Paradigm}&{Reference}   \\  
			\hline
			\multicolumn{1}{|c|}{\multirow{5}{*}{Job Scheduling}} &
			\multicolumn{1}{c|}   {GA-PSO} & {makespan, cost, execution time}&{CC} & {\cite{Ahmad124}}  \\ \cline{2-5}
			\multicolumn{1}{|c|}{}                                                 &
			\multicolumn{1}{c|}  {GA-ACO} & {execution time} &{CC}& {\cite{Liu226}} \\ \cline{2-5}
			\multicolumn{1}{|c|}{} &
			\multicolumn{1}{c|}  {GA-ACO} & {energy consumption, makespan, resource utilization} &{MCC}& {\cite{RASHIDI230}}   \\ \cline{2-5}
			\multicolumn{1}{|c|}{} &
            \multicolumn{1}{c|}  {ACO-CSA} & {energy consumption, makespan} &{CC}& {\cite{N.Moganarangan192}}   \\ 
			\hline
			\multicolumn{1}{|c|}{\multirow{1}{*}{Resource Allocation}} &
			\multicolumn{1}{c|} {RL-ACO} & {throughput}&{MEC} &{\cite{Vimal290}}  \\
			\hline
			\multicolumn{1}{|c|}{\multirow{1}{*}{Task Offloading}} &
			\multicolumn{1}{c|}  {NN-PSO} & {security and privacy}&{CC} & {\cite{Ahmad165}} \\ 
			\hline
			\multicolumn{1}{|c|}{\multirow{2}{*}{Joint Issues}} &
	  	    \multicolumn{1}{c|}  {GA-PSO} &  {energy consumption}& {MEC} &  {\cite{Guo270}} \\ \cline{2-5}
	  	    \multicolumn{1}{|c|}{} &
	  	    \multicolumn{1}{c|}  {DRL-PSO-FL} &  {energy consumption} & {MEC}&  {\cite{Jiang271}}  \\ 
			\hline
		\end{tabular}
		\label{tab-HS}
	\end{center}
\end{table*}

Chen \emph{et al.} \cite{Yuanfang166} proposed a DL-based model to detect security threats in MEC. This model uses unsupervised learning and location information to improve the detection process, which can detect malicious applications at the edge of a cellular network. Numerical results demonstrate that the proposed model outperforms softmax regression, decision tree, support vector machine, and random forest.
Rani and Pounambal \cite{Rani218} proposed DL-based dynamic task offloading in mobile cloudlet, which considers energy consumption and execution time as objective metrics. The task computed on cloudlet or cloud server is divided into subtasks. Experimental results show that the proposed algorithm outperforms cloudlet-based dynamic task offloading in terms of energy consumption and completion time. 
Dinh \emph{et al.} \cite{Dinh213} proposed a RL-based computation offloading scheme in MEC to reduce energy consumption. They studied multi-user multi-edge-node computation offloading problem, and formulated it as a non-cooperative game where each user maximizes its own utility. Experimental results show that the proposed scheme outperforms local processing and random assignment.
Meng \emph{et al.} \cite{Meng214} proposed a DRL-based task offloading algorithm to minimize the mean slowdown of tasks and energy consumption of MEC. A new reward function is designed to optimize the tradeoff between average slowdown and average energy consumption. Numerical results reveal that the proposed algorithm performs better than the baseline algorithms such as all in MEC server, all in mobile device, and random in terms of average energy consumption and average slowdown.

Chen \emph{et al.} \cite{Xianfu188} presented two DRL algorithms for optimizing computation offloading performance in virtual EC, where the stochastic computation offloading problem is formulated as an MDP. Numerical results show that the proposed algorithms outperform mobile execution, server execution, and greedy execution in terms of computation offloading performance.
Zhang \emph{et al.} \cite{Zhang224} proposed a DRL-based task offloading scheme for vehicular edge computing networks (VECNs), in which the central cloud server is considered as a backup server due to its powerful computation capacities. The task offloading problem is formulated as a processing time minimization problem with delay constraints. 
Ning \emph{et al.} \cite{Ning241} studied DRL for intelligent Internet of vehicles. They constructed an offloading framework consisting of three layers (i.e., cloudlet, RodeSide Units, and fog nodes) to minimize the total energy consumption under the delay constraint. The proposed method outperforms the baseline algorithms in terms of average energy consumption.
Ranadheera \emph{et al.} \cite{Shermila187} proposed a computation offloading approach based on game theory and RL in MEC to reduce energy consumption. 

Wang \emph{et al.} \cite{Wang272} proposed a DRL-based method for computation offloading in MEC. The offloading problem in MEC is formulated as an MDP and the S2S neural network is designed to represent the offloading policy. Simulation results show that the proposed method performs better than two heuristic baselines in terms of latency, and can obtain nearly optimal results while having polynomial time complexity.
Chen \emph{et al.} \cite{Chen275} proposed a DRL-based approach for performance optimization in MEC. They considered MEC for a representative mobile user in an ultra-dense network, where one of multiple base stations can be selected for computation offloading. They modeled an optimal computation offloading policy as an MDP and developed a deep Q-network-based strategic computation offloading algorithm to learn the optimal policy without having any priori knowledge of the dynamic statistics.
Chen and  Wang \cite{Chen279} proposed a DRL-based approach for decentralized computation offloading in MEC to minimize the long-term average computation cost in terms of power consumption and buffering delay. They adopted continuous action space-based DRL to learn efficient computation offloading policies independently at each mobile user. Experimental results demonstrate that the proposed approach outperforms conventional deep Q-network-based discrete power control strategy and some other greedy strategies in terms of computation cost.

\subsubsection{LBS for Joint Issues}
Huang \emph{et al.} \cite{Huang238} proposed distributed DL-based offloading in MEC. They formulated joint offloading decision and bandwidth allocation as a mixed-integer programming problem. Multiple parallel deep NNs are used to generate offloading decisions. Numerical results reveal that the proposed algorithm outperforms deep Q-network in terms of total cost and energy consumption.
Huang \emph{et al.} \cite{LiangHuang182} further studied the same joint issue in multi-user MEC by adopting the idea of deep Q-network. Experimental results show that the proposed algorithm performs better than the MUMTO algorithm in \cite{chen2016joint} in terms of overall cost, energy consumption, and delay.
Liu \emph{et al.} \cite{Liu222} formulated offloading and resource allocation in VECNs as a semi-Markov process, by considering stochastic vehicle traffic, dynamic computation requests, and time-varying communication conditions. Two RL methods, i.e., a Q-learning based method and a DRL method, are designed to get the optimal policies for computation offloading and resource allocation. 
Ning \emph{et al.} \cite{Ning223} constructed an intelligent offloading system for VECNs by using DRL. They used Markov chains to model communication and computation states. To improve users' quality of experience, task scheduling and resource allocation are formulated as a joint optimization problem. To schedule offloading requests and allocate network resources, this paper designs a two-sided matching scheme and a DRL approach. Experimental results show that the proposed method outperforms Q-learning, a greedy method, local computing, and deep Q-network in terms of quality of experience and execution time.
Xiao \emph{et al.} \cite{Liang169} applied RL to provide secure offloading to the edge nodes against jamming attacks in MEC. Lightweight authentication and secure collaborative cashing schemes are presented for securing data privacy. Simulation results show that the proposed RL-based secure solution can effectively enhance the security and user privacy of MEC and protect it against various smart attacks with low overhead.
Li \emph{et al.} \cite{Li273} proposed a DRL-based computation offloading and resource allocation scheme for MEC. In order to minimize the sum cost of delay and energy consumption for all user equipments in MEC, they jointly optimized the offloading decision and computing resource allocation. Simulation results reveal that the proposed scheme performs better than other baselines.
Huang \emph{et al.} \cite{Huang274} proposed a DRL-based method for online computation offloading in wireless powered MEC networks to maximize the weighted sum computation rate with binary computation offloading. They optimally adapted task offloading decisions and wireless resource allocations to the time-varying wireless channel conditions. Simulation results demonstrate that the proposed algorithm achieves similar near-optimal performance as existing benchmark methods but reduces the CPU execution latency by more than an order of magnitude.

\subsection{HS in CC and EC}
This section discusses the works on the use of HS for solving issues in CC and EC, which are summarized in Table \ref{tab-HS}.

\subsubsection{HS for Job Scheduling}
Manasrah and Ali \cite{Ahmad124} proposed a hybrid GA-PSO algorithm for workflow task scheduling in CC. It outperforms GA, PSO, and some other CI techniques in terms of total execution time of the workflow task. 
Liu \emph{et al.} \cite{Liu226} proposed a hybrid GA-ACO task scheduling algorithm in CC, which takes advantage of fast convergence from ACO and global search ability of GA. Simulation results show that it outperforms GA and ACO in terms of task execution time.
Rashidi and Sharifian \cite{RASHIDI230} presented a hybrid GA-ACO algorithm for task scheduling in MCC. Experimental results show that it outperforms queue-based round robin, queue-based random, and queue-based weighted round robin assignment algorithms in terms of average completion time, total energy consumption of mobile devices, and the number of dropped tasks.
Moganarangan \emph{et al.} \cite{N.Moganarangan192} combined ACO with CSA to address job scheduling in CC. The proposed algorithm performs better than ACO in terms of energy consumption and makespan.

\subsubsection{HS for Resource Allocation}
Vimal \emph{et al.} \cite{Vimal290} proposed a hybridization of RL and multi-objective ACO to enhance resource allocation in MEC for industrial IoT. The proposed algorithm allocates resources accurately and optimally for users in MEC. Experimental results show that the proposed algorithm outperforms GA and BA in terms of throughput. 

\subsubsection{HS for Task Offloading}
Saljoughi \emph{et al.} \cite{Ahmad165} presented a hybrid NN-PSO algorithm to detect intrusions and attacks in CC. For extraction of the optimal weights of NN, the weights of NN are optimized by using PSO. The proposed method obtains better outcomes than the simple NN on NSL-KDD and KDD-CUP databases.

\subsubsection{HS for Joint Issues}
Guo \emph{et al.} \cite{Guo270} studied the energy-efficient computation offloading management scheme based on hybrid GA-PSO in MEC with small cell networks. They jointly optimized computation offloading, spectrum, power, and computation resource to minimize the energy consumption of all users' equipments. They presented the computation offloading model and formulated the problem as a mixed-integer nonlinear programming problem. Simulation results show that the proposed algorithm performs better than other baseline algorithms.
Jiang \emph{et al.} \cite{Jiang271} presented a hybrid  DL-PSO-FL algorithm for online offloading to minimize the energy consumption of users' equipments in a hybrid MEC network. In the proposed algorithm, FL is used to locate the ground vehicles and UAVs, PSO is used to solve the mixed-integer nonlinear programming problem and provides high-quality samples to DNN, and DL is applied to make the task admission and resource allocation decision in real time. Simulation results reveal that the proposed algorithm reduces the CPU time by more than several orders of magnitude.

\begin{figure}
	\centering
	\includegraphics[width=0.50\textwidth]{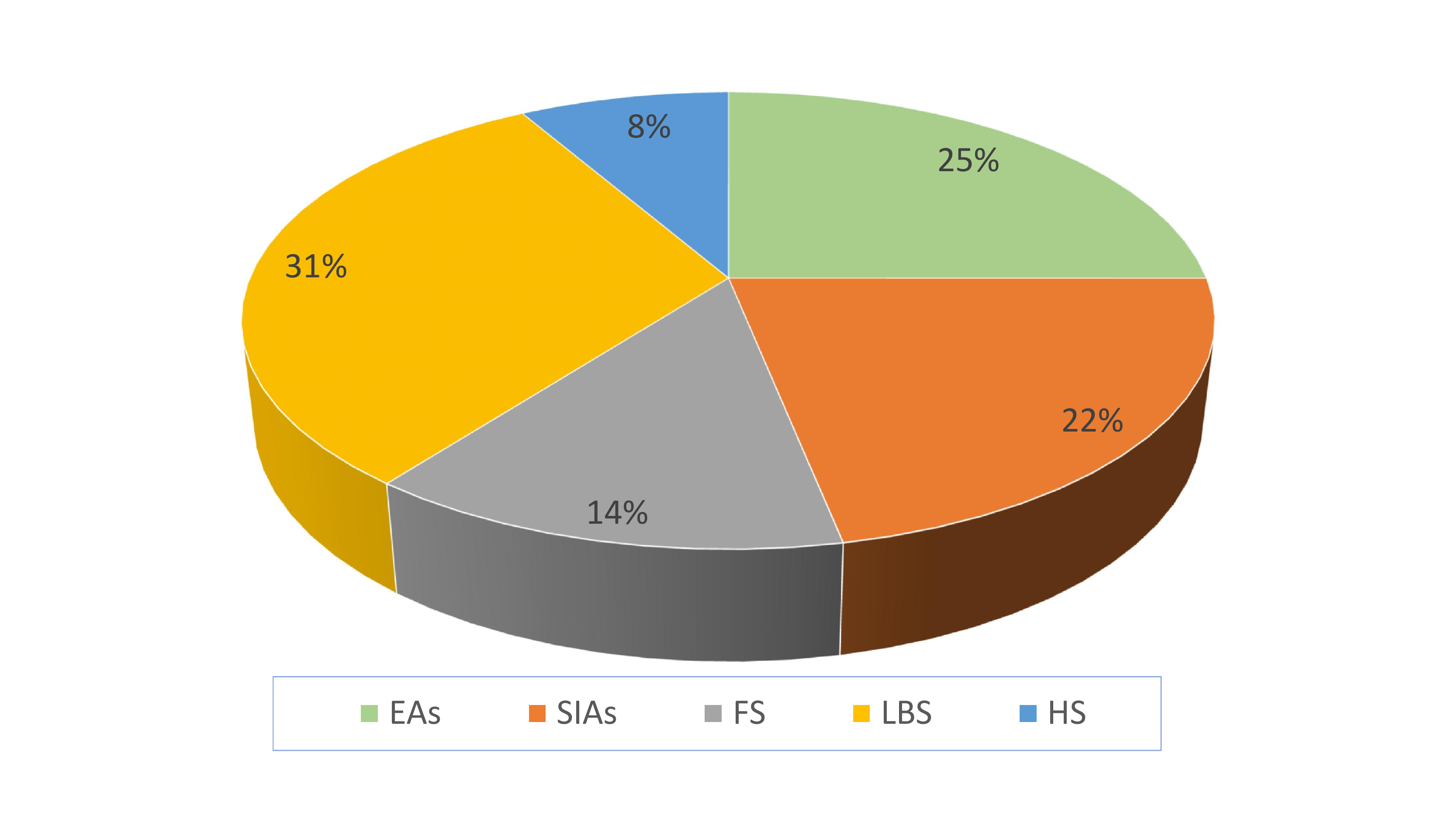}
	\caption{Overall percentage distributions of CI techniques for addressing issues in CC and EC.}
	\label{CI-CC-DIST}
\end{figure}

\begin{figure}
	\centering
	\includegraphics[width=0.53\textwidth]{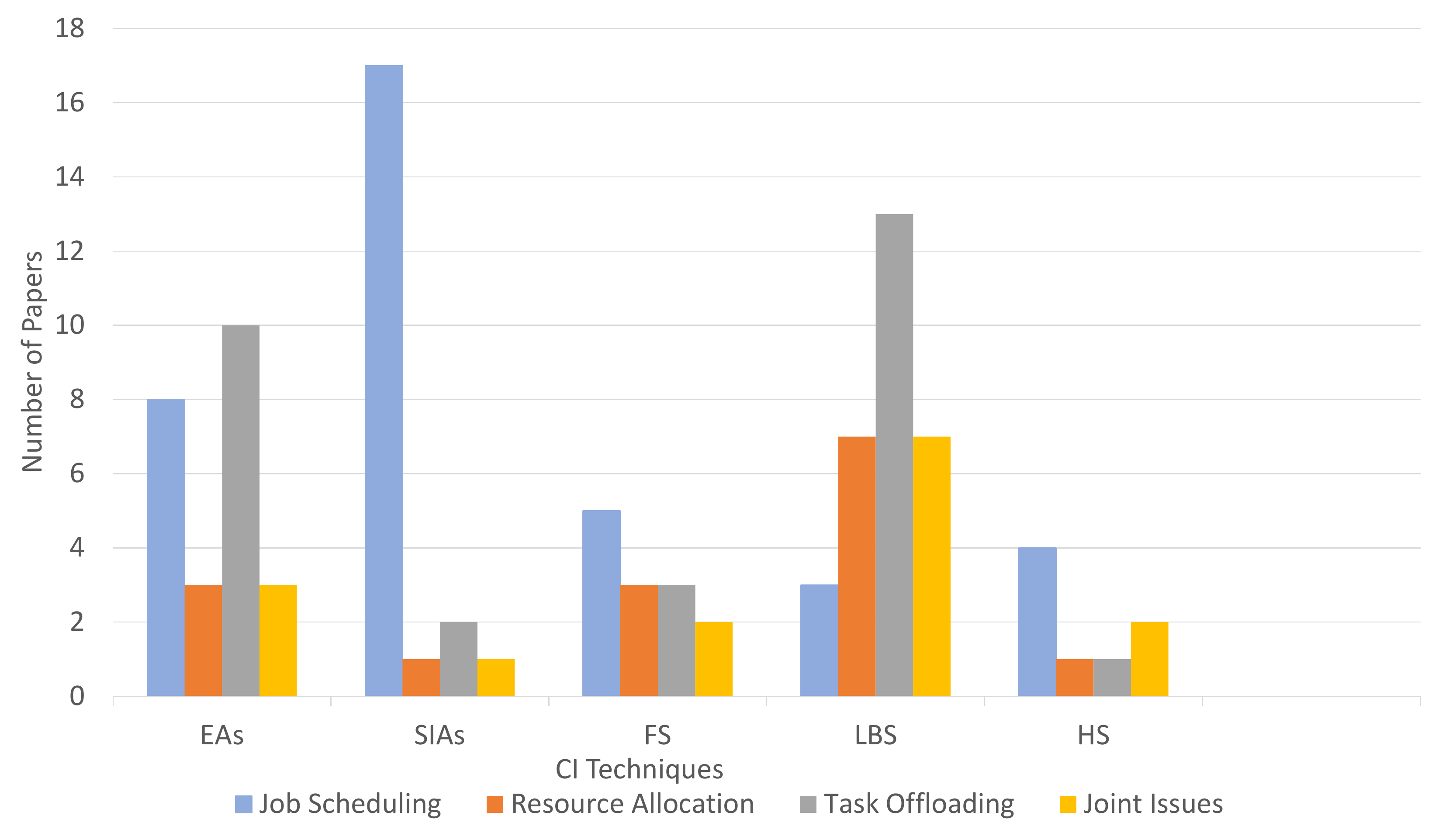}
	\caption{Suitability of CI techniques for addressing issues in CC and EC.}
	\label{CI-ISSUE-DIST}
\end{figure}

\begin{figure}[t]
	\centering
	\includegraphics[width=0.44\textwidth]{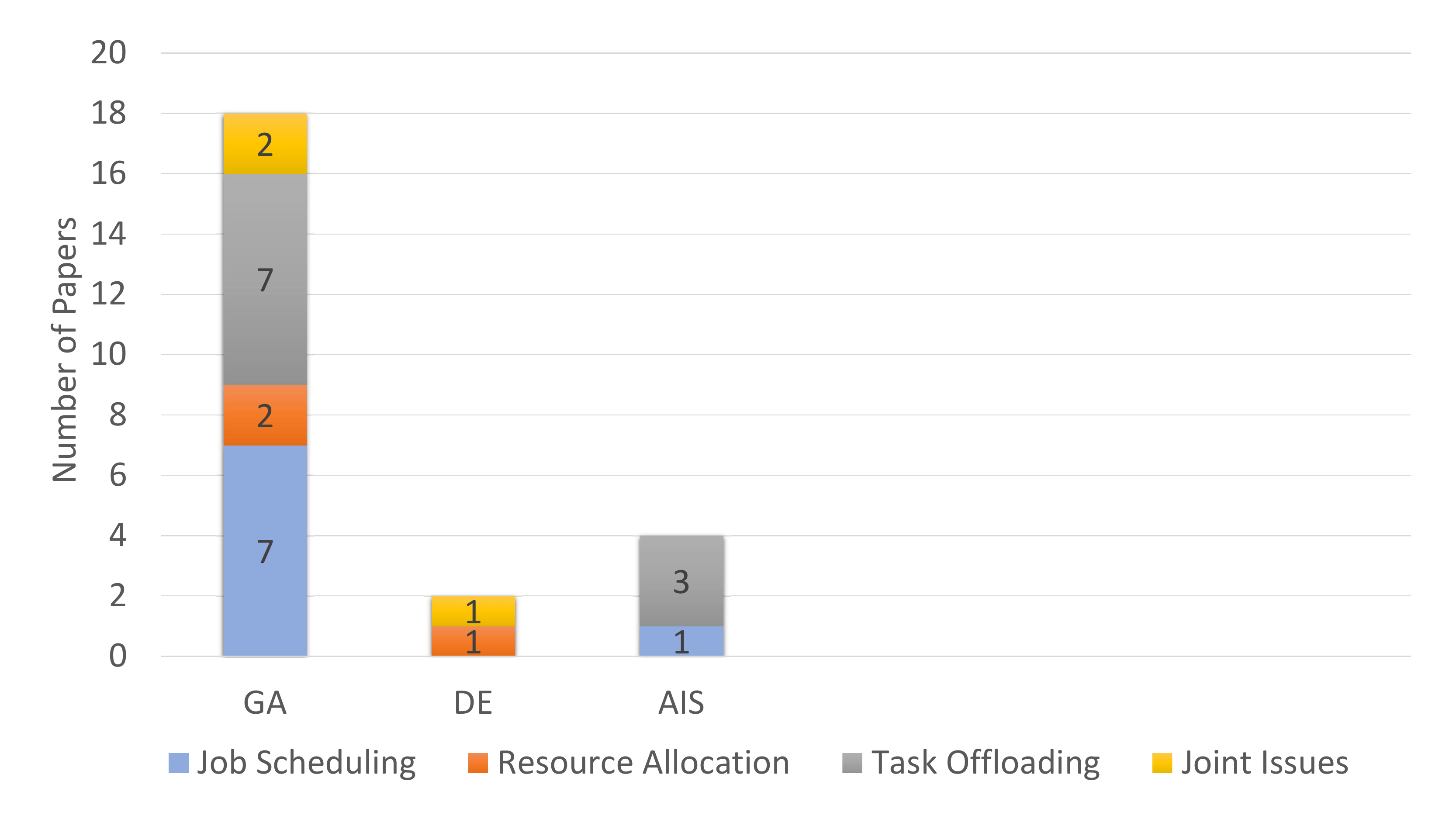}
	\caption{Distribution of EAs in CC and EC.}
	\label{EAs-CC-EC}
\end{figure}

\begin{figure}[t]
	\centering
	\includegraphics[width=0.44\textwidth]{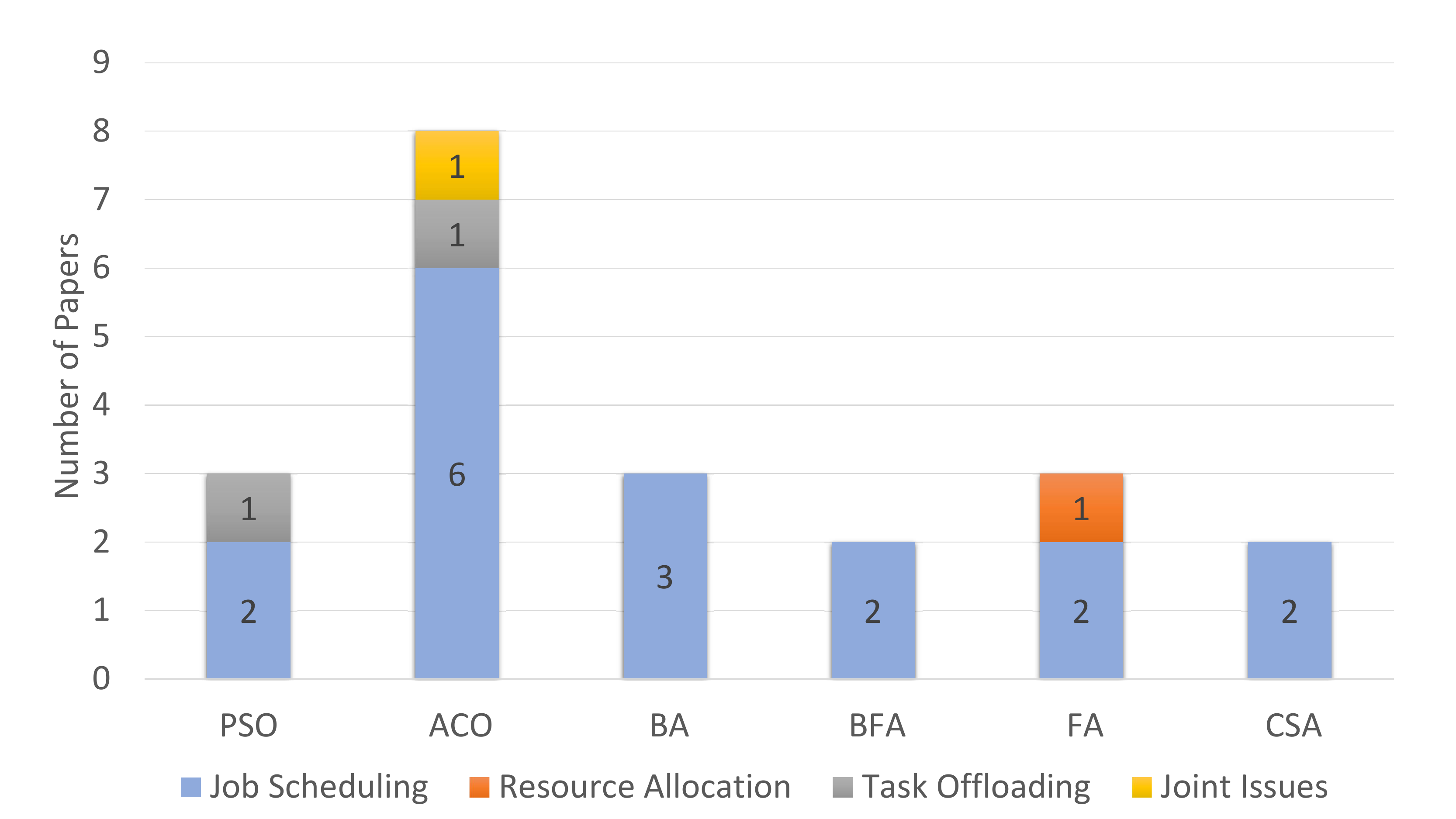}
	\caption{Distribution of SIAs in CC and EC.}
	\label{SIAs-CC-EC}
\end{figure}

\begin{figure}[t]
	\centering
	\includegraphics[width=0.44\textwidth]{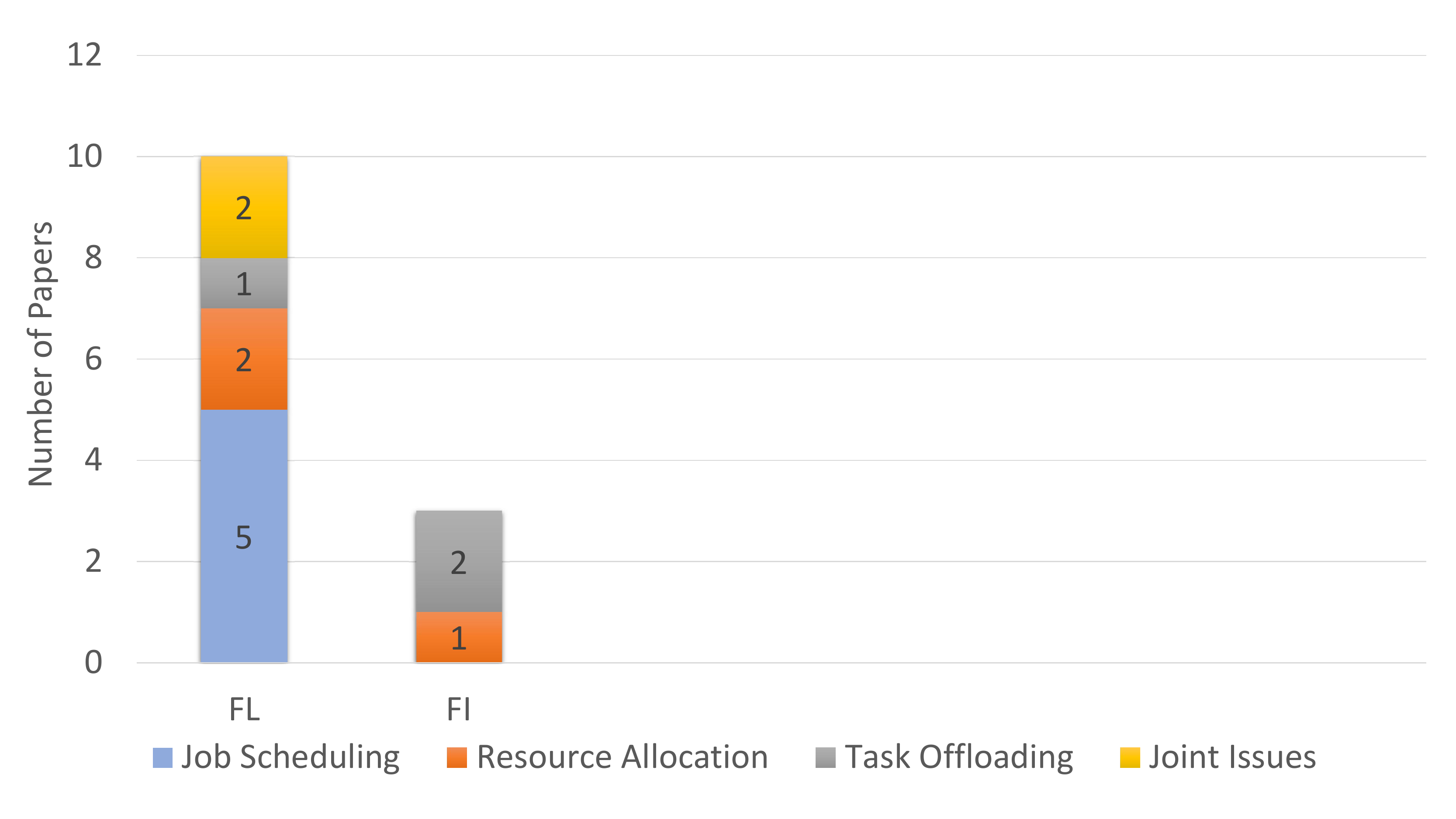}
	\caption{Distribution of FS in CC and EC.}
	\label{FS-CC-EC}
\end{figure}

\begin{figure}[t]
	\centering
	\includegraphics[width=0.44\textwidth]{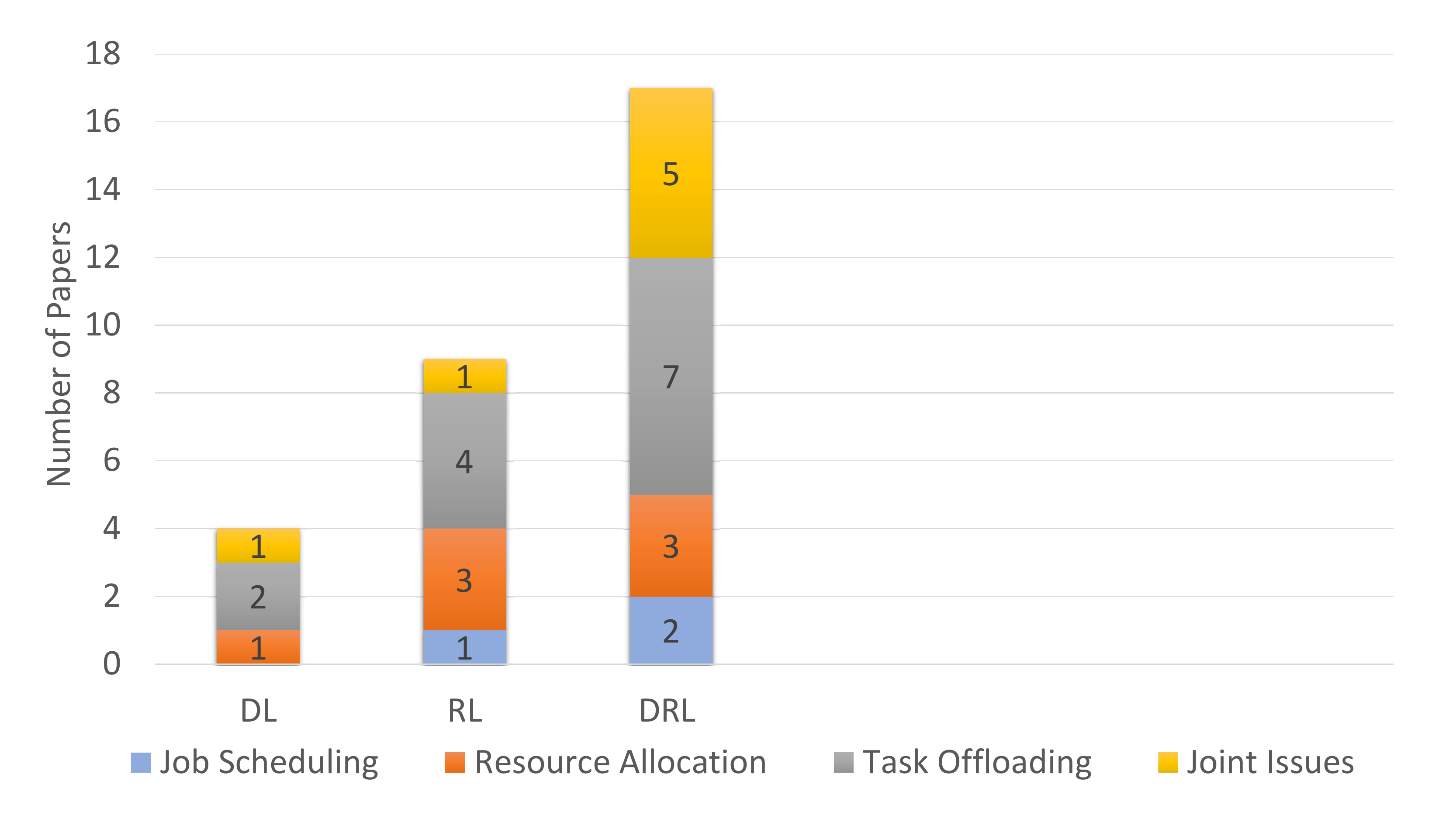}
	\caption{Distribution of LBS in CC and EC.}
	\label{LS-CC-EC}
\end{figure}

\section{Discussion and Future Research Trends}\label{Discussion-future}
Recent literature shows that researchers have paid much attention to the innovative use of CI techniques to address issues in CC and EC, such as job scheduling, resource allocation, task offloading, and joint issues. However, there are still open challenges for researchers to tackle. This section provides the statistics which reflect the status-quo of CI for CC and EC, and points out future research trends.

\subsection{Discussion}
Fig. \ref{CI-CC-DIST} presents the overall percentage distributions of CI techniques mentioned in Fig. \ref{FIG-CI} for solving issues in CC and EC. Fig. \ref{CI-ISSUE-DIST} presents all issues, CI techniques, and the number of papers found for each issue in CC and EC. Furthermore, we elaborate on the main research findings derived from Section \ref{App-CI-CC-EC}. 

\subsubsection{EAs for CC and EC}
From Fig. \ref{CI-CC-DIST}, EAs are the second most commonly used techniques among all types of CI techniques. Most of the works found in this survey are on the use of EAs for job scheduling and task offloading as shown in Fig. \ref{EAs-CC-EC}. Furthermore, the most frequently used EA in CC and EC is GA as shown in Fig. \ref{EAs-CC-EC}. Fig. \ref{EAs-CC-EC} also presents the distribution of EAs, which indicates the suitability of each kind of EAs for each issue in CC and EC.

\subsubsection{SIAs for CC and EC}
SIAs have been used in 22\% papers compared with other CI techniques as shown in Fig. \ref{CI-CC-DIST}. We can observe from Fig. \ref{SIAs-CC-EC} that job scheduling is the hottest issue addressed by SIAs and ACO is the most frequently used SIA in CC and EC. Fig. \ref{SIAs-CC-EC} also shows the distribution of SIAs in the context of issues in CC and EC.

\subsubsection{FS for CC and EC}
It can be seen from Fig. \ref{CI-CC-DIST} that 14\% papers employ FS to address issues in CC and EC. FS has been frequently used for job scheduling, resource allocation, and task offloading as shown in Fig. \ref{FS-CC-EC}. Fig. \ref{FS-CC-EC} presents the distribution of FS in the context of issues in CC and EC.

\subsubsection{LBS for CC and EC}
LBS is the most commonly studied CI technique in CC and EC as shown in Fig. \ref{CI-CC-DIST}. In particular, it has been used more intensively for task offloading, resource allocation, and joint issues in CC and EC from Fig. \ref{LS-CC-EC}. In addition, DRL is the most frequently used LBS as shown in Fig. \ref{LS-CC-EC}. Fig. \ref{LS-CC-EC} gives the distribution of LBS in CC and EC.

\subsubsection{HS for CC and EC}
HS has not been utilized frequently (only 8\% papers) in CC and EC as depicted in Fig. \ref{CI-CC-DIST}. Specifically, half of  papers related to HS studies on job scheduling as shown in Fig. \ref{CI-ISSUE-DIST}.

\textbf{Remark:} Based on the above discussion, we would like to give more details to understand what kinds of CI techniques are good at solving what kinds of issues:
\begin{itemize}
	\item GA is the best choice for solving job scheduling and task offloading among all EAs \cite{ZIXUE183, Binh245, Zarina118, Guo270}. The reasons are twofold: 1) it is the most popular and frequently used EA paradigm; and 2) it has various crossover and mutation operators that can deal with different optimization problems, such as discrete and continuous optimization problems.
	\item ACO is the best SIA for job scheduling among all the issues studied in this paper \cite{Zehua130, Wei231, Wang229}, due to the following reasons: 1) it can consider/incorporate multiple scheduling targets/metrics, such as load balancing, energy consumption, makespan, and cost; 2) each variable in the solution obtained by ACO is generated one by one; thus, jobs can be assigned to VMs one by one to optimize the scheduling process; and 3) it can ensure the fast convergence and good performance by balancing the exploration of new solutions and exploitation of accumulated experience about the problem \cite{X.F.LIU193}.
	\item Compared with FI, FL is better for solving job scheduling, resource allocation, and joint issues \cite{Anindita107, Srinivas123}, since it is flexible, easy to understand, compatible with the uncertainty of CC and EC parameters as well as users' behaviors, and can deal with imprecise data and complex problems with several variables.
	\item DRL performs the best among all LBS for resource allocation, task offloading, and joint issues \cite{Li273, Nguyen160, Xiong291, Ning223}. The superiority of DRL can be summarized as follows: 1) it can automatically adapt and customize itself according to users' requirements \cite{Wang278}; 2) it can discover/learn new knowledge from large databases; 3) it can develop models that are difficult and expensive to be designed manually due to the requirements of specific skills; and 4) it has a strong ability to handle complex problems by efficiently learning from experiences.
\end{itemize}

\subsection{Future Research Trends}

\subsubsection{From CC and EC Perspective}
\begin{itemize}
\item Many-Metric Formulation: More metrics need to be considered at the same time in CC and EC to create the trade-off among them. For example, in the future, we may consider the trade-off among latency, cost, energy consumption, security, and privacy to satisfy QoS of users.
\item Edge Node Allocation: Due to the distributed nature of EC, edge clouds that offer services across diverse geolocation and regions are difficult to be allocated \cite{WANG269}. In the future, efficient service discovery protocols are needed to design, such that users can identify and locate the relevant service providers to meet their demands.
\item Real-Time Optimization: For many edge application scenarios, the service environments are of high dynamics \cite{Canali252} and it is hard to correctly foresee future events. Thus, it would require the remarkable capabilities of online edge resource orchestration and provisioning to continuously handle massive dynamic workloads and tasks. In the future, real-time resource allocation is required to fulfill dynamic task demands.
\item Decentralized Trust: The open nature of EC leads to the decentralized trust, e.g., services provided by different edge entities must be secured and trustworthy. Thus, efficient security mechanisms are required to ensure users' authentication, data integrity, and mutual platform verification for EC \cite{Zhou247}. Also, novel secure routing schemes and trust network topologies are critical for EC. In addition, end devices would generate a large volume of data at the network edge, which can be privacy-sensitive since they may contain users' location data, health status, personal activities records or other sensitive information \cite{Jianbing46}. Therefore, feasible paradigms are needed to secure data sharing to minimize the privacy leakage.
\item Hardware Constraints in EC: Due to hardware constraints, unlike CC, EC cannot support heavy-weight software \cite{Jianbing46}. Big data analysis and data warehousing will never be feasible with existing EC because of the increasing number of duplication of small software in market. Therefore, users are facing difficulties in looking for a trusted edge provider among edge providers with a lack of standardized framework. Thus, developing software and hardware for handling computation offloading from the central cloud is a critical issue to be addressed.
\item Migration of EC Applications: Migration of EC applications among different edge clouds is a challenging issue that can help to balance loads or accommodate users' movements and minimize the end-to-end latency of users \cite{Chen246}. In the future, efficient techniques are required based on system measurements and experiments to handle the migration challenges.
\item Interoperability and Collaboration of CC and EC: Thanks to EC, users are able to process latency-sensitive applications at the network edge \cite{Naha280}. However, handling an increasing number of multiple services is still a big challenge. Some tasks may be redirected to the central  clouds for further processing. Thus, it is important to develop effective architectures and efficient algorithms to facilitate the above collaboration.
\item Heterogeneity: The future edge and cloud networks may present a huge level of heterogeneity, i.e., we may come across various kinds of users, ranging from smartphones, smartwatches, sensor nodes, and intelligent cameras \cite{Zhang295}. This may cause the instability of the networks and, therefore, efficient solutions are highly required \cite{Mithun287}.
\item Edge Intelligence: There is a growing trend to process data at edge clouds because of privacy and other concerns \cite{Zhou247}. However, due to the limited resources in edge clouds, some computational-intensive tasks such as machine learning models may not be trained at edge clouds without any proper modification. Therefore, in the future, it is important to study how to execute or train machine learning models in edge clouds. Federated learning has been proposed to address this issue but more attempts are expected in the future.
\item Blockchain-Assisted CC and EC: Blockchain has attracted much attention from both academia and industry \cite{Yuan293, Choo294}. It is also interesting to design the blockchain-based networks for CC and EC. Three aspects can be considered here: 1) blockchain can be applied to enhance the privacy in the communication and computing resource sharing; 2) blockchain can be very useful in terms of improving the price and security of CC and EC; and 3) as blockchain may need huge computing resources to run, CC and EC can provide computing resources to blockchain-related tasks or applications.
\end{itemize}

\subsubsection{From CI Technique Perspective}
\begin{itemize}
 \item Execution Time: Most CI techniques like EAs need a long time for searching the optimal solution in CC and EC. In the future, we may find a way to decrease the execution time.
\item Convergence: Sometimes, it is not easy to prove the convergence in CI techniques. However, in CC and EC, it is important to have an algorithm with convergence guarantee. Therefore, a future trend is to design some CI techniques with good convergence performance.
\item Multi/Many-Objective Optimization: Two or more objectives need to be addressed in CC and EC to meet the service level agreements of users. Therefore, a future trend may be to design multi/many-objective CI techniques to address multi/many-objective optimization problems in CC and EC.
\item Constraint-Handling Issue: Since most CI techniques cannot handle constraints effectively, we can incorporate some constraint-handling techniques into CI techniques to address constrained optimization problems in CC and EC.
\item CI Techniques for Security and Privacy: Directly sharing data among multiple edge nodes may run a high risk of privacy leakage. Therefore, federated learning paradigms can be applied to secure data sharing by training distributed data such that the original data sets can be kept in their source nodes and the edge AI model parameters can be shared among different nodes. 
\item AI Edge Models: Instead of utilizing the existing resource-intensive AI models in CC and EC, we can design a resource-aware edge AI model. For example, methods like CI techniques can be adopted to efficiently search over the AI model design parameter space by taking into account the impact of hardware resource constraints on the performance metrics such as execution latency, security, and energy overhead.
\item CI Techniques for Migration of EC Applications: Sharing and migrating applications among edge clouds is a challenging task, as we have to consider the load balancing and reduce the latency for all tasks. LBS and FS may be suitable candidates for migration of EC applications.
\item CI Techniques for Interoperability and Collaboration of CC and EC: Multiple latency-sensitive requests may arrive at the same time. Therefore, fast CI techniques (such as LBS and FS) along with prediction algorithms are potential for the collaboration between the central and edge clouds.
\item Hybrid Versions: Hybrid CI techniques have shown great potential in solving complex optimization problems; however, few papers have studied them for addressing issues in CC and EC. A future trend may be to design hybrid CI techniques. For instance, FS is suitable for job scheduling while LBS is competitive for task offloading. In the future, we may combine FS and LBS to address joint issues. Similarly, more hybrid CI techniques can be developed to deal with energy consumption, completion time, and security in CC and EC.
\end{itemize}

\section{Conclusion}\label{Conc-CI-CC}
In this survey paper, we reviewed the applications of CI techniques to four critical issues in CC and EC: job scheduling, resource allocation, task offloading, and joint issues. We commenced with rudimentary concepts of CC and EC along with critical issues and metrics, and then focused on five categories of CI techniques used in CC and EC: EAs, SIAs, FS, LBS, and HS. Subsequently, diverse designed approaches relying on CI techniques were reviewed in the context of CC and EC. In addition, the statistics about the status-quo of CI for CC and EC were provided based on the works collected in this survey paper. We found that LBS was intensively used in CC and EC, followed by EAs and SIAs. However, FS and HS were not fully utilized in CC and EC. Finally, we pointed out some challenges and future research trends in CI, CC, and EC.

\bibliographystyle{IEEEtran}
\bibliography{AsimBib}

\end{document}